\documentclass[%
reprint, 
superscriptaddress, 
amsmath, amssymb, 
aps, 
prl,
longbibliography]{revtex4-1}

\usepackage{graphicx}
\usepackage{amsfonts}    
\usepackage{verbatim}   
\usepackage{color}      
\usepackage{subfigure}  
\usepackage{hyperref}   
\usepackage{booktabs}
\usepackage{threeparttable}
\usepackage{dcolumn}
\usepackage{multirow}
\usepackage{physics}

\begin{document}

\title{Unbalanced-basis-misalignment tolerant measurement-device-independent quantum key distribution}

	\author{Feng-Yu Lu}\email{These authors contribute equally to this work}
	\author{Ze-Hao Wang}\email{These authors contribute equally to this work}
	\author{Zhen-Qiang Yin}\email{yinzq@ustc.edu.cn}
	\author{Shuang Wang}\email{wshuang@ustc.edu.cn}
	\affiliation{CAS Key Laboratory of Quantum Information, University of Science and Technology of China, Hefei, Anhui 230026, P. R. China}
	\affiliation{CAS Center for Excellence in Quantum Information and Quantum Physics, University of Science and Technology of China, Hefei, Anhui 230026, P. R. China}
	\affiliation{State Key Laboratory of Cryptology, P. O. Box 5159, Beijing 100878, P. R. China}
	\author{Rong Wang}
  \affiliation{Department of Physics, University of Hong Kong, Pokfulam Road, Hong Kong SAR, China}
	\affiliation{CAS Key Laboratory of Quantum Information, University of Science and Technology of China, Hefei, Anhui 230026, P. R. China}
	\affiliation{CAS Center for Excellence in Quantum Information and Quantum Physics, University of Science and Technology of China, Hefei, Anhui 230026, P. R. China}
    \author{Guan-Jie Fan-Yuan}
    \author{Xiao-Juan Huang}
    \author{De-Yong He}
	\author{Wei Chen}
	\author{Zhen Zhou}
	\author{Guang-Can Guo}
	\author{Zheng-Fu Han}
	\affiliation{CAS Key Laboratory of Quantum Information, University of Science and Technology of China, Hefei, Anhui 230026, P. R. China}
	\affiliation{CAS Center for Excellence in Quantum Information and Quantum Physics, University of Science and Technology of China, Hefei, Anhui 230026, P. R. China}
	\affiliation{State Key Laboratory of Cryptology, P. O. Box 5159, Beijing 100878, P. R. China}


\date{\today}

\begin{abstract}
Measurement-device-independent quantum key distribution (MDIQKD) is a revolutionary protocol since it is physically immune to all attacks on the detection side. However, the protocol still keeps the strict assumptions on the source side that the four BB84-states must be perfectly prepared to ensure security. Some protocols release part of the assumptions in the encoding system to keep the practical security, but the performance would be dramatically reduced. In this work, we present a MDIQKD protocol that requires less knowledge of encoding system to combat the troublesome modulation errors and fluctuations. We have also experimentally demonstrated the protocol. The result indicates the high-performance and good security for its practical applications. Besides, its robustness and flexibility exhibit a good value for complex scenarios such as the QKD networks.

\end{abstract}

\pacs{Valid PACS appear here}
\maketitle


\section{Introduction}

Quantum key distribution(QKD)\cite{BB84} allows two remote users, called Alice and Bob, to share secret random keys because of the information-theoretical security guaranteed by principles of quantum physics \cite{lo1999unconditional,shor2000simple,scarani2009security,Rennersecurity,pirandola2020advances}, even if there is an eavesdropper, Eve. Unfortunately, practical QKD systems still suffer from troublesome attacks \cite{makarov2006effects,zhao2008quantum,lydersen2010hacking,pang2020hacking,huang2020laser} rooted in the gaps between theoretical models and practical setups. The device-independent quantum key distribution (DIQKD) \cite{acin2007device} is intrinsically immune to all side-channel attacks but not practically usable due to its exorbitant demand for the detection efficiency and channel loss. As an alternative, some protocols \cite{braunstein2012side,lo2012measurement} are proposed to remove all side-channels on the vulnerable detection side. The measurement-device-independent quantum key distribution (MDIQKD) \cite{lo2012measurement} is naturally immunes to all detection-side-channel attacks \cite{makarov2006effects,zhao2008quantum,lydersen2010hacking} while maintaining a comparable performance \cite{tang2014measurement,yin2016measurement} to the regular prepare-and-measure QKD systems. The MDIQKD has received extensive attention since it is a perfect balance between security and practicality.

The good properties of the MDIQKD have aroused widespread interest in recent years \cite{liu2013experimental,da2013proof,tang2014measurement,pirandola2015high,yin2016measurement,tang2016measurement,comandar2016quantum,roberts2017experimental,liu2019experimental,semenenko2020chip,wei2020high,woodward2021gigahertz}. However, a fly in the ointment is that the security of the MDIQKD relies on an assumption that Alice and Bob can fully control their source side and perfectly prepare the four ideal BB84 states \cite{BB84,ferenczi2012security}, which still impedes the unconditional security and hinders the practical application. In practical systems, the preparation of the four states usually relies on the accurate modulation \cite{yin2016measurement,tang2014measurement,wang2015phase,liu2018polarization,boaron2018secure,zhou2021reference}, which is still a great challenge due to the unbalanced path-loss of interferormeters\cite{ferenczi2012security}, the limited precision of modulation signals, insufficient system bandwidth, errors in calibration, and impacts of environments \cite{yoshino2018quantum,roberts2018patterning,lu2021intensity,zhang2020state}. When the prepared states in MDIQKD are non-ideal BB84 states, the users can not accurately estimate the information leakage anymore since the important security assumption that "the density matrices of Z and X base should be indistinguishable" is broken. Indeed, the encoders can be calibrated in advance, but the cumbersome calibrations would unavoidably increase the experimental difficulty. Besides, it may introduce additional loopholes \cite{jain2011device}.

Some security proofs \cite{ferenczi2012security,wang2013three,yin2013measurement,yin2014mismatched,hwang2017improved,zhou2020experimental,tamaki2014loss,tang2016experimental,zeng2020symmetry,jin2021reference,coles2016numerical,winick2018reliable,primaatmaja2019versatile,bourassa2020loss} successfully remove part of the demands in the coding system and greatly improve the practical security of the MDIQKD. However, some of them over-pessimistically estimate the information-leakage bound so that the protocol performance would dramatically decrease with the misalignment \cite{ferenczi2012security,yin2013measurement,yin2014mismatched,hwang2017improved,zhou2020experimental}. Some other proofs require additional informations about the maximum misalignment \cite{tamaki2014loss,tang2016experimental,bourassa2020loss} or previously characterizing the states \cite{coles2016numerical,winick2018reliable,primaatmaja2019versatile}, which may increase the experimental complexity.
To promote the practical application, in this work, we propose a MDIQKD whose source-side restrictions are greatly reduced.
In this protocol, the users prepare a general Z basis and a "simplified X basis" that could be "unbalanced", which is common scenarios in practical time-bin phase coding systems. The protocol has an invariable high-performance against the X-basis imbalances, and maintains its security even if the phase reference frame between the two users is misaligned \cite{laing2010reference,yin2014reference}, the prepared states of the "X basis" are mixed states, and the prepared states of the Z basis are also misaligned.

In this work, we first introduce our protocol with the ideal single-photon sources. After that, an improved analysis is proposed for the scenarios that the states in the Z basis are not pure. To make our protocol practically useful with the existing weak coherent sources, the decoy-state method \cite{hwang2003quantum,wang2005beating,lo2005decoy} are also designed and a joint-study method \cite{yu2015statistical,zhou2016making,lu2020efficient} against the statistical fluctuation \cite{ma2012statistical,tomamichel2012tight,curty2014finite,lim2014concise,lim2021security} is proposed. Finally, we built a time-bin phase coding MDIQKD system to experimentally demonstrate this protocol in the non-asymptotic cases. The 25 MHz system ran continuously for accumulating sufficient data in several scenarios with different imbalances. The results indicate that the protocol can tolerate large imbalances and confirm its security and feasibility for simplifying the experimental system, which would be promising for practical application and network scenarios. The details of security proofs against the collective attack and the parameter estimations can be found in Supplemental Material.

\section{Single-photon protocol}

In the original MDIQKD, the four BB84 states, in another words, $\ket{0}$ and $\ket{1}$ as the Z basis, $\ket{+} = \frac{\sqrt{2}}{2}\left(\ket{0} + \ket{1}\right)$ and $\ket{-} = \frac{\sqrt{2}}{2}\left(\ket{0} - \ket{1}\right)$ as the X basis, are required \cite{lo2012measurement}. Our protocol reduces the demands so that Alice and Bob can prepare an unbalanced "X basis" that satisfies 
\begin{equation}
\begin{aligned}
\label{state_relations}
&\ket{\varphi_2} = c_0 \ket{0} + c_1 \ket{1},\  \ket{\varphi_3} = c_0 \ket{0} - c_1 \ket{1},\\
&\ket{\varphi'_2} = c'_0  \ket{0} + c'_1e^{i\theta} \ket{1},\ \ket{\varphi'_3} = c'_0 \ket{0} - c'_1 e^{i\theta} \ket{1},\\
\end{aligned}
\end{equation}	
where $\ket{\varphi_2}$ and $\ket{\varphi_3}$ ($\ket{\varphi'_2}$ and $\ket{\varphi'_3}$) correspond to Alice's (Bob's) original $\ket{+}$ and $\ket{-}$ respectively, the $\theta$ denotes the phase-reference frame misalignment \cite{laing2010reference,yin2014reference} between two users. We use $\beta$ ($\beta'$) to describe the imbalance misalignment of Alice (Bob), and the positive real number coefficients $c_0=\cos(45^\circ + \beta)$ and $c_1=\sin(45^\circ + \beta)$ ($c'_0=\cos(45^\circ + \beta')$ and $c'_1=\sin(45^\circ + \beta')$) respectively, where $\beta\in (-45^\circ,45^\circ)$. The protocol has an invariant high performance with different $\beta$, and maintains its security when the reference-frame misalignment $\theta \neq 0$ and the states in "X basis" are not pure due to modulation fluctuations. As a matter of fact, the prepared mixed states in "X basis" can be regarded as uncharacterized $\ket{\varphi_2}$ and $\ket{\varphi_3}$ in Eq.(\ref{state_relations}). The detail of proof can be found in Supplemental Material.

In each turn, Alice (Bob) randomly selects one of the four states and sends it to untrusted Charlie to perform a measurement. Charlie publicly announces if he has measured a successful event. If Charlie is honest, he should do the Bell state measurement (BSM) and announce the $\ket{\psi^{-}} = \frac{\sqrt{2}}{2}\left( \ket{01}- e^{i\theta} \ket{10} \right)$. Charlie may not perform the required measurement because he is unreliable and $\theta$ is an unknown value, but the property of the MDIQKD guarantees the protocol security and an appropriate $\theta$ can maximize the secret key rate.

\noindent
\fbox{%
\label{box1}
  \parbox{8.3cm}{%
      \textbf{Box.1: Protocol procedure}\\ \\
      \textbf{1.Preparation:} In each turn, Alice (Bob) randomly prepares quantum state $\ket{0}$, $\ket{1}$, $\ket{\varphi_2}$ and $\ket{\varphi_3}$ ($\ket{0}$, $\ket{1}$, $\ket{\varphi'_2}$ and $\ket{\varphi'_3}$) and sends it to the untrusted measurement unit Charlie. We define that the code basis (the Z basis) is selected when the user prepares $\ket{0}$ or $\ket{1}$ and the test basis (the unbalanced X basis) is selected if Alice (Bob) prepares $\ket{\varphi_2}$ or $\ket{\varphi_3}$ ($\ket{\varphi'_2}$ or $\ket{\varphi'_3}$).\\ \\
      \textbf{2.Measurement:} Charlie projects his received pulse-pair to the Bell-state $\ket{\psi^- }= \frac{1}{\sqrt{2}}(\ket{0}\ket{1} + e^{-i\theta}\ket{1}\ket{0})$ and publicly announces success or failure in each turn. Alice and Bob generate theit raw key bits according to Charlie's announcement\\ \\
      \textbf{3.Sifting:} After the above trial has been repeated enough times, Alice and Bob publicly announce their basis for each turn. If both of them select the code basis, they maintain the raw key bit. Otherwise the raw key bit is discarded. The remained raw key bits are named sifted key bits.\\ \\
	    \textbf{4.Parameter estimation:} By sacrificing some of the data for public discussion, users estimate the single-photon yield $q_{nm}$. According to the $q_{nm}$, the users estimate the bit and phase error rate.\\ \\
      \textbf{5.Error correction and security amplification:} According to the above estimated parameters, Alice and Bob perform the error correction and security amplification to generate the secret key bits.
  }%
}\\

According to the announcement of success or failure, Alice and Bob record or discard the data. When the users have accumulated sufficient data, they sacrifice some of the data for public discussion to estimate the $q_{nm}$, which is defined as the yield when Alice codes $\ket{\varphi_n}$ and Bob codes $\ket{\varphi'_m}$. Especially, when both Alice and Bob select the Z basis, the raw key bits are generated according to the data. Other data are used in parameter estimations. The positive real coefficients $c_0$, $c_1$, $c'_0$, $c'_1$ can be accurately calculated by
\begin{equation}
\begin{aligned}
&c_0=\sqrt{\frac{q_{10}q_{T1}-q_{11}q_{T0}}{q_{01}q_{10}-q_{00}q_{11}}},c_1=\sqrt{\frac{q_{01}q_{T0}-q_{00}q_{T1}}{q_{01}q_{10}-q_{11}q_{00}}},\\
&c_0'=\sqrt{\frac{q_{01}q_{1T}-q_{11}q_{0T}}{q_{01}q_{10}-q_{00}q_{11}}},c_1'=\sqrt{\frac{q_{10}q_{0T}-q_{00}q_{1T}}{q_{01}q_{10}-q_{11}q_{00}}}.
\end{aligned}
\end{equation}
where $q_{Tm}=\frac{q_{2m}+q_{3m}}{2}$, $q_{nT}=\frac{q_{n2}+q_{n3}}{2}$ ($n,m\in\{0,1\}$).

The information leakage is bounded by $ H(e_p)$, where the $H(x) = -x\log_2(x) - (1-x)\log_2(1-x)$ denotes the binary Shannon entropy function and the $e_p$ is the phase error rate of the $Z$ basis, which is described as
\begin{equation}
\begin{aligned}
\label{phase_error}
e_p  = \frac{1}{2} - \frac{\left(q_{23} + q_{32} \right) - \left( q_{22} + q_{33} \right)}{8\left( q_{00} + q_{01} + q_{10} + q_{11} \right)c_0c'_0c_1c'_1},
\end{aligned}
\end{equation}
which is invariant as if the relation in Eq.(\ref{state_relations}) is met. In another word, Eve's information can be precisely bounded even if the "X basis" is unbalanced. We note that the $e_p$ would jump to 0.5 in several extreme cases that the four quantum states are not different from each other (One may see Supplemental Material for more details.). The secret key rate (SKR) is described by the Shor-Preskill formula \cite{shor2000simple}
\begin{equation}
\begin{aligned}
\label{key_rate_single_photon}
R = q_{C}\big(1 - H(e_p) - H(e_{b})\big),
\end{aligned}
\end{equation}
where $q_{C} = (q_{00}+q_{01}+q_{10}+q_{11})/4$ denotes the yield when Alice and Bob both select the Z basis, and the $e_{b} = (q_{00}+q_{11})/(q_{00}+q_{01}+q_{10}+q_{11})$ denotes the bit error rate when both of the users select the Z basis.

\begin{figure}[htbp]
	\includegraphics[width=9cm]{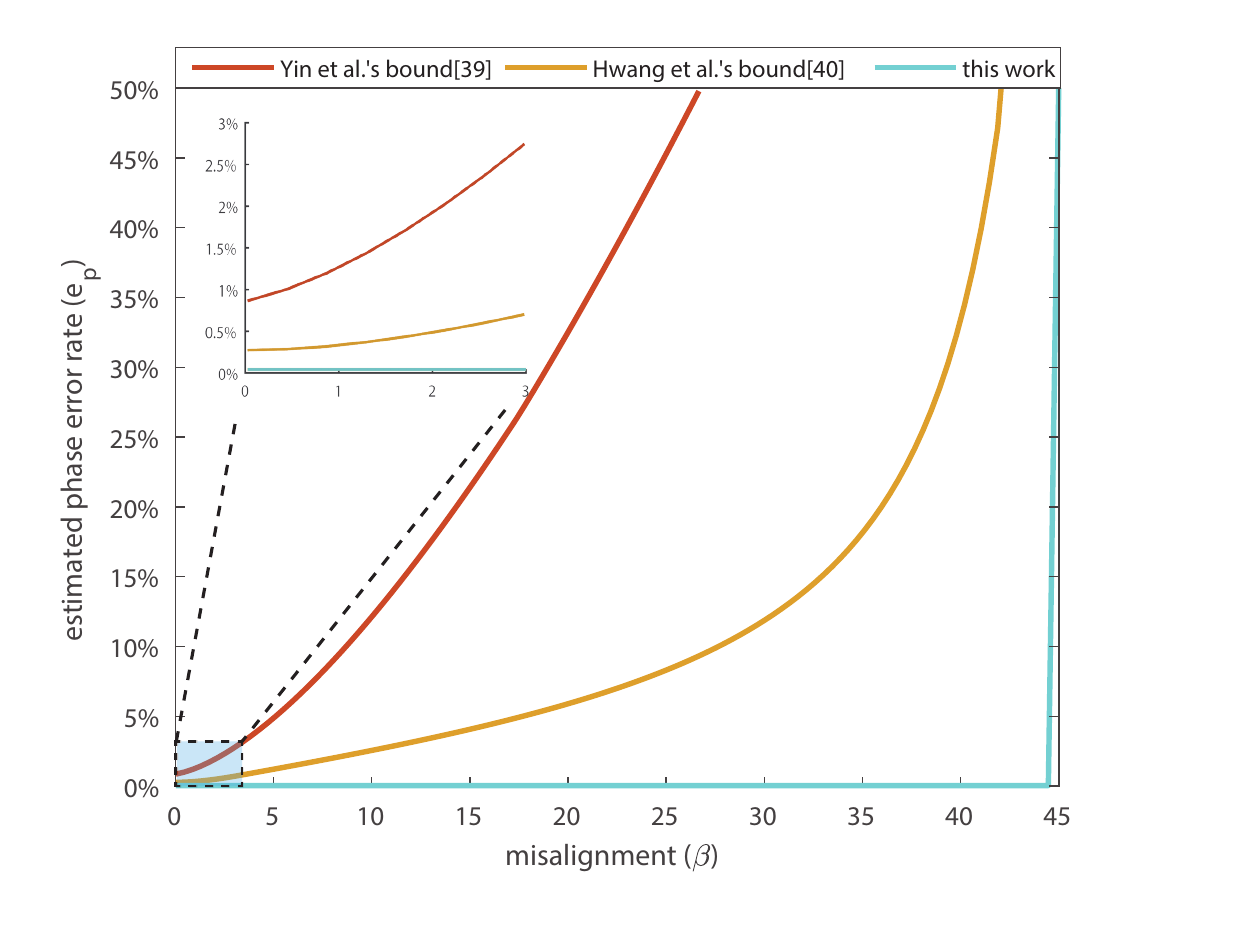}
	\caption{\label{Fig_inf_decoy} The estimated phase error for several single-photon MDIQKDs versus the imbalance misalignment $\beta$. The performance of the uncharacterized qubits MDIQKDs reduced with the increasing $\beta$. In contrast, the performance of our protocol is nearly a constant value before approaching the extreme point that $\beta = 45^\circ$. The simulation parameters are $P_d = 3\times10^{-6}$ (dark count rate), $P_\eta = 20\%$ (detection efficiency), and the transmission loss is 20 dB.   }
\end{figure}

To demonstrate the property against the imbalance misalignment, we simulated our single-photon protocol and several ideal single-photon uncharacterized-qubits MDIQKDs \cite{yin2014mismatched,hwang2017improved}. As illustrated in Fig. \ref{Fig_inf_decoy}, the performance of the uncharacterized-qubits MDIQKDs decay with the increasing $\beta$ ($\beta'$), in contrast, the performance of our protocol is invariant.

The protocol procedure is described in the Box.1\ref{box1}

\section{protocol with mixed states}

In practical systems, the prepared states could be mixed states due to the modulation fluctuations. Our protocol can maintain its security in these mixed-state cases and we prove this property in two steps.

\textbf{test basis:}

We first consider the scenario that the test basis is not pure. Here we only introduce our key idea and the details of proof are in the Supplemental Material. Our key idea is proving that the mixed states in our test basis equal to the uncharacterized states $\ket{\varphi_2}$ and $\ket{\varphi_3}$ with Eve's operation in the quantum channel.

We take Alice as an example, due to the modulation fluctuations, the misalignment $\beta$ is not a fixed value but satisify the random distribution with probability density function $P(\beta)$. 
The test basis can be described as mixed states
\begin{equation}
\begin{aligned}
& \rho_2 
 = \frac{1}{2}(\mathbb{I} + \hat{S}_X\sigma^\theta_X + \hat{S}_Z\sigma_Z ),\\
&\rho_3 
= \frac{1}{2}(\mathbb{I} - \hat{S}_X\sigma^\theta_X + \hat{S}_Z\sigma_Z )
\end{aligned}
\end{equation}
where $\sigma_Z$ and $\sigma^\theta_X$ are Pauli matrices, $\mathbb{I}$ is the identity matrix, $\hat{S}_Z$ ($\hat{S}_X$) denotes the Z (X) component of the prepared state in the Bloch sphere. The prepared state is a pure state if $\hat{S}^2_Z + \hat{S}^2_X = 1$ and a mixed state if $\hat{S}^2_Z + \hat{S}^2_X < 1$.

Defining Eve's operation 
\begin{equation}
\varepsilon(\rho) = \lambda_0 \mathbb{I}\rho\mathbb{I} + \lambda_1 \sigma_Z \rho \sigma_Z, 
\end{equation}
which satisfying
\begin{equation}
\begin{aligned}
\left \{
\begin{array}{lc}
\lambda_0 + \lambda_1 = 1,\\
(\lambda_0 - \lambda_1) \sqrt{1 - {\hat{S}_Z}^2} = \hat{S}_X.
\end{array}
\right.
\end{aligned}
\end{equation}
The $\varepsilon$ can be intuitively regarded as Eve doing nothing with the probability $\lambda_0$ and performing the $\sigma_Z$-operation with the probability $\lambda_1$. We can find that the $\varepsilon$ satisfies
\begin{equation}
\begin{aligned}
& \ket{0}\bra{0} = \varepsilon(\ket{0}\bra{0}), \ \ \ket{1}\bra{1} = \varepsilon(\ket{1}\bra{1}),\\
& \rho_2 = \varepsilon \left(\ket{\varphi_2}\bra{\varphi_2}\right), \ \   \rho_3 = \varepsilon \left(\ket{\varphi_3}\bra{\varphi_3}\right).
 \end{aligned}
\end{equation}
So the protocol security in the case that the test-basis has modulation fluctuation equals to the security in the case that the users prepare pure-states $\ket{0}$, $\ket{1}$ and uncharacterized $\ket{\varphi_2}$, $\ket{\varphi_3}$ while Eve performs $\varepsilon$ in the quantum channel. Noting that the operations in the quantum channel have nothing to do with security, we can claim that our protocol is secure in this scenario.

\textbf{code basis:}

The code basis is usually well-aligned in practical time-bin phase coding systems so that the pure-state protocol is enough for most cases. However, if the Z basis is out of control due to the modulation error and fluctuation, we should consider an improved analysis. Here we still only introduce our key idea and leave the details in our Supplement Material.

In this scenario, when Alice wants to prepare $\ket{0}$ ($\ket{1}$), she actually randomly prepares $\cos( \beta_0 )\ket{0} + e^{i\theta}\sin( \beta_0 )\ket{1}$ or $\cos( \beta_0 )\ket{0} - e^{i\theta}\sin( \beta_0 )\ket{1}$ ( $\sin( \beta_1 )\ket{0} + e^{i\theta}\cos( \beta_1 )\ket{1}$ or $\sin( \beta_1 )\ket{0} - e^{i\theta}\cos( \beta_1 )\ket{1}$ ), where the randomness of "$+$" and "$-$" derive from Alice's randomly modulation of $0$ or $\pi$ relative phase between the two time bins when selecting the code basis. Besides, because of the modulation fluctuations, the misalignment $\beta_0$ and $\beta_1$ are not fixed value but random distribution with probability density functions $P_0(\beta_0)$ and $P_1(\beta_1)$ respectively. In other words, Alice actually prepares mixed states 
\begin{equation}
\begin{aligned}
& \rho_0  = (1 - \xi)\ket{0}\bra{0} + \xi \ket{1}\bra{1},\\
& \rho_1  = \zeta\ket{0}\bra{0} + (1-\zeta) \ket{1}\bra{1},
\end{aligned}
\end{equation}
The $\xi$ and $\zeta$ are defined as the misalignment errors of Alice's $\ket{0}$ and $\ket{1}$ respectively.

Similarly, Bob's $\ket{0}$ and $\ket{1}$ are changed to
\begin{equation}
\begin{aligned}
& \rho'_0  = (1 - \xi')\ket{0}\bra{0} + \xi' \ket{1}\bra{1},\\
& \rho'_1  = \zeta'\ket{0}\bra{0} + (1-\zeta') \ket{1}\bra{1},
\end{aligned}
\end{equation}
respectively, and the $\xi'$ and $\zeta'$ are misalignment errors of Bob's $\ket{0}$ and $\ket{1}$ respectively.

In the mixed state scenario, the observable values are the mixed-state yields $y_{nm}$ rather than $q_{nm}$. Our key idea is bounding the pure-state yields $q_{nm}$ by the observed $y_{nm}$ and the lower and upper bounds of the Z basis misalignments, namely, the $\upsilon^L$ and $\upsilon^U$ where $\upsilon \in \{\xi,\zeta,\xi',\zeta' \}$. Indeed, the misalignment errors are unknown values but their lower and upper bounds can be estimated by previously calibration or be monitored by inserting a local single-photon detector (similar to the monitor SPD-Moni in Fig.\ref{experiment_setup}). As far as the $q_{nm}$ is bounded, the phase error $e_p$ can be estimated similar to the pure-state case.

We have also analyzed the decoy-state method and the statistical fluctuation for the mixed-state scenarios. The details are also in our Supplemental Material.

\section{protocol with decoy-state method} 

The above protocol is based on ideal single-photon sources, which are still not practically available. So we propose a four-intensity decoy-state method to connect the theory with practice. In this section we only analyze the asymptotic case that the data size is infinite, and the non-asymptotic that considers the statistical fluctuation \cite{ma2012statistical} is introduced in Supplemental Material. 

In our method, Alice (Bob) prepares a phase randomized weak coherent pulse and randomly selects an intensity $l$ ($r$) from a pre-decided set $\{\mu, \nu, \omega, o\}$ ($\{\mu', \nu', \omega', o\}$), where the $\mu$ ($\mu'$) is defined as signal state and the others are defined as decoy states. Especially, the $o=0$ is also named vacuum state. If the signal state is selected, Alice (Bob) only selects the code basis. Else if other intensities are selected, they prepare the four quantum states just like the single-photon protocol. Alice and Bob record their data according to Charlie's announcement and sacrifice some of them for estimating the gains. They should publicly announce which intensity is selected. If both of them select the signal state, the recorded data would become the raw key bit. In other cases, they would announce which state is selected. According to their announcement, they can calculate out the gains $Q_{nm}^{lr}$ where the subscript $nm$ denotes Alice and Bob prepare $\ket{\varphi_n}$ and $\ket{\varphi'_m}$ ($\ket{\varphi_0}=\ket{\varphi'_0}=\ket{0}$, $\ket{\varphi_1}=\ket{\varphi'_1}=\ket{1}$) respectively, and the superscript $lr$ denotes Alice and Bob select intensity $l$ and $r$ respectively. With these $Q_{nm}^{lr}$, Alice and Bob can bound the single-photon yields tightly \cite{hwang2003quantum,wang2005beating,lo2005decoy,wang2013three,yu2013three,yu2015statistical,zhou2016making} so that the SKR can be described as  
\begin{equation}
\begin{aligned}
\label{key_rate_WCP}
R = p_\mu p_{\mu'}\left[ a_1^\mu b_1^{\mu} \underline{q}_{C}\left( 1 - H(\overline{e}_p) \right) - Q_{C}^{\mu\mu}fH(E_{C}^{\mu\mu}) \right],
\end{aligned}
\end{equation}
where the $p_\mu$ and $p'_\mu$ denote the probability of selecting the signal state, the $a_1^{\mu} = \mu e^{-\mu}$ ($b_1^{\mu} = \mu' e^{-\mu'}$) is the Poisson distribution probability for sending a single-photon state, the $Q_{C}^{\mu\mu}$ and $E_{C}^{\mu\mu}$ are the observed gain and the observed error rate of the signal state pulse-pairs respectively, $f=1.16$ is the error correction efficiency, and the underline and overline are lower and upper bound respectively.

\begin{figure}[htbp]
	\includegraphics[width=9cm]{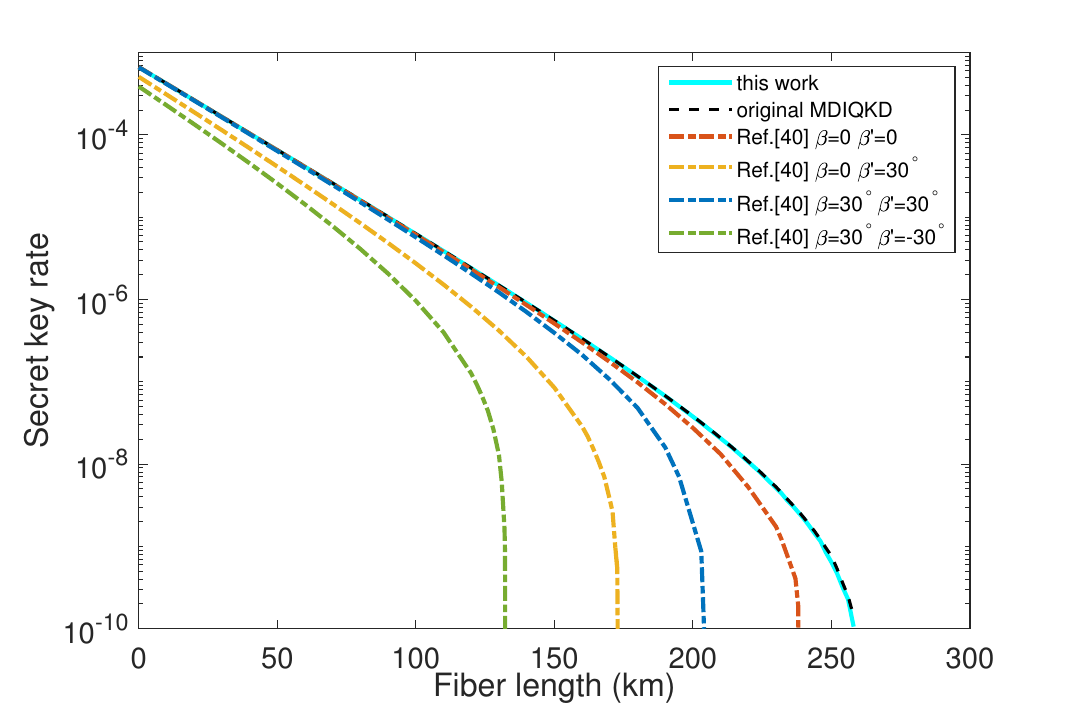}
	\caption{\label{Fig_asymp_R_L} The SKR as a function of fiber length of different decoy-state MDIQKDs in asymptotic cases. We simulated our protocol in many different combinations of $\beta$ and $\beta'$ and find that their asymptotic SKR are nearly indistinguishable in this figure so we only use a cyan solid line to represent our protocol with different $\beta$ and $\beta'$. As a contrast, we simulate the asymptotic SKR of the original MDIQKD (black dash line) without any misalignment ($\beta = \beta' =0$) . We can find that in asymprotic case, our protocol with the imbalance misalignment has the same performance compared with ideal original MDIQKD. We also simulated the state-of-the-art uncharacterized qubits MDIQKD \cite{hwang2017improved,zhou2020experimental}(dot-dash line) with different $\beta$ and $\beta'$ and use lines in different colors to denote different imbalances. red: $\beta = \beta' =0$, yellow: $\beta = 0, \beta' =30^\circ$, blue: $\beta = 30^\circ, \beta' =30^\circ$, green: $\beta = 30^\circ, \beta' =-30^\circ$. The simulation parameters are $P_d = 3\times10^{-6}$, $P_\eta = 20\%$, and the fiber loss is 0.2 dB/km.  }
\end{figure}

\begin{figure}[htbp]
\includegraphics[width=9cm]{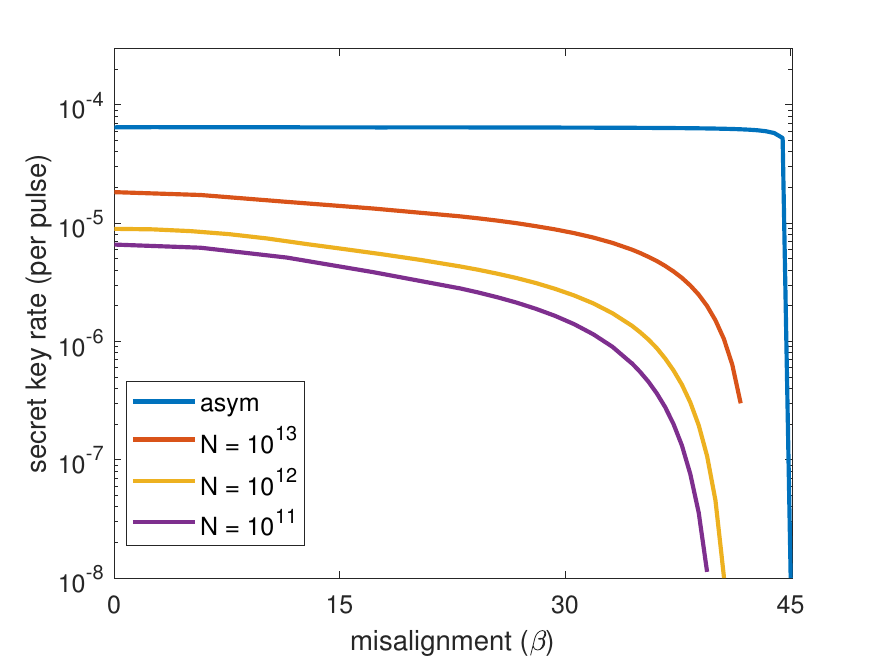} 
\caption{\label{Fig_R_beta} The secret key rate as a function of the misalignment $\beta$ in the "decoy-state cases". The blue line denotes the asymptotic case that the data size is infinite. The red, yellow, and purple lines denote the SKR with the data size of $10^{13}$, $10^{12}$, $10^{11}$ respectively. The simulation parameters are $P_d = 3\times10^{-6}$, $P_\eta = 20\%$, fiber length is 50 km and the fiber loss is 0.2 dB/km, failure probability $\epsilon$ in using chernoff bound \cite{zhang2017improved} is $5.73\times10^{-7}$.}
\end{figure}

Figure \ref{Fig_asymp_R_L} shows the asymptotic SKR of the four-intensity decoy-state method. When the X basis is unbalanced, our protocol maintains an invariant result while other protocols may not work. Especially, the plots of our protocol with different imbalances are nearly overlapping with the original MDIQKD with perfect coding, which indicates higher robustness and better practicality of our protocol. Fig. \ref{Fig_R_beta} shows the simulation of the unbalanced-basis-misalignment tolerance. We can find that the asymptotic secret key rate is nearly invariant and the non-asymptotic secret key rate decreases slowly, which indicates that the asymptotic $e_p$ is near invariant before closing to the extreme point and the non-asymptotic $e_p$ is slightly affected by the unbalanced-basis-misalignment $\beta$ (the details of the non-asymptotic cases are introduced in Supplemental Material).

It worth noting that the above analysis is suitable for the pure-state scenario. The analysis for the mixed-state scenario is a little different, we would introduce its detail in the Supplemental Material.
\\

\section{Experimental demonstration}

We experimentally demonstrated our protocol in the non-asymptotic cases (the parameter estimation for the non-asymptotic cases is in Supplemental Material.) As illustrated in Fig. \ref{experiment_setup}, our experiment setup consists of two identical legitimate users and an untrusted relay. 
Each of the legitimate users employs a continuous-wave laser (Wavelength References Clarity-NLL-1542-HP) that is frequency locked to a molecular absorption line of 1542.38 nm center wavelength.
Then the continuous-wave lasers are chopped \cite{wang2015phase,wang2017measurement,liu2018polarization,zhou2020experimental} into a 500 ps temporal width pulse sequence with a 25 MHz repetition by the intensity modulator IM-1, which is driven by a homemade narrow-pulse generator. The phase modulator PM-1 next to the IM-1 actively randomizes the phase of the pulses to avoid imperfect-source attacks \cite{sun2012partially,xu2013practical}.
 \begin{figure*}[htbp]
	\includegraphics[width=18cm]{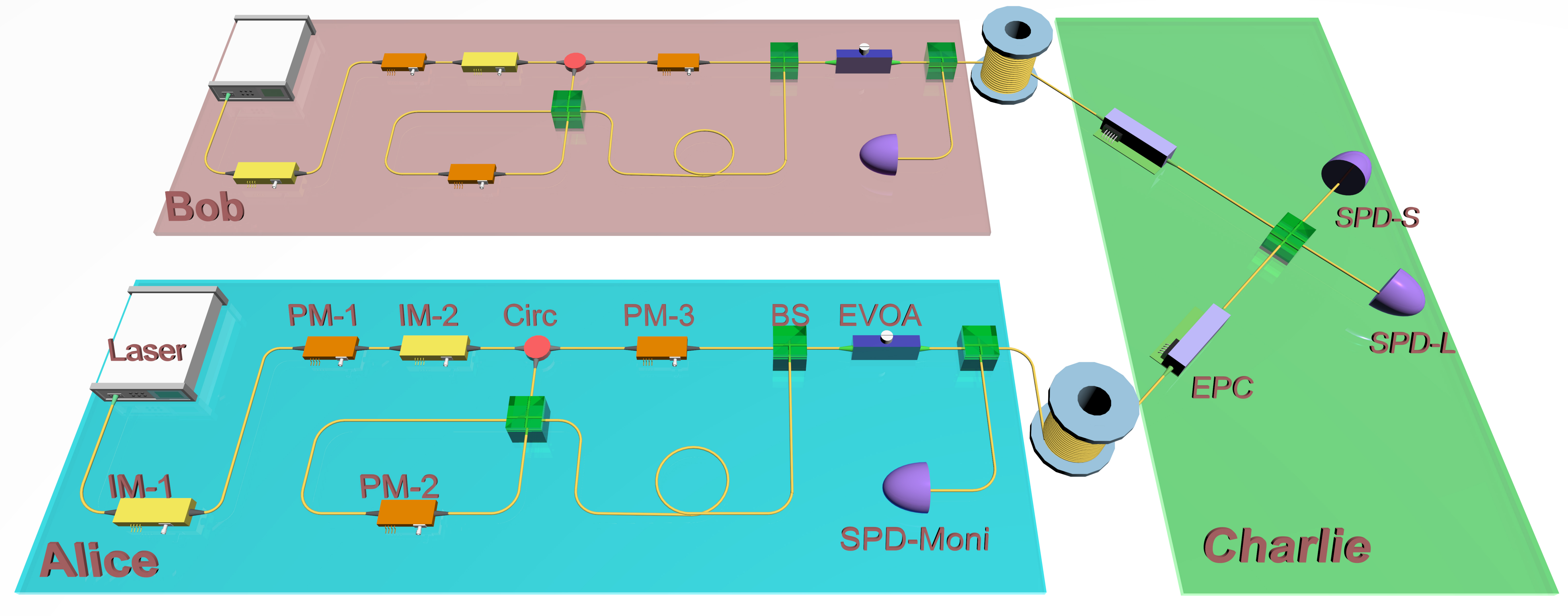}
	\caption{\label{experiment_setup} Diagram of our experimental system. Laser, continuous-wave laser; IM, intensity modulator; PM, phase modulator; Circ, circulator; BS, beam splitter; EVOA, electric variable optical attenuator; EPC, electric polarization controller; SPD, single-photon detector.}
\end{figure*}
After that, the phase randomized coherent pulses are fed into the intensity modulator IM-2 to randomly modulate four different intensities for our decoy-state method. The IM-2 is driven by a homemade two-bit digital to analog converter (DAC) with a 25 Mbps bitrate, and the digital pseudo random numbers come from an FPGA-based driver board (PXIe-6547, National Instruments). Besides, when the vacuum state is selected, the IM-1 and IM-2 eliminate the pulse jointly to achieve an over 50 dB extinction ratio.

Following the IM-2, a Sagnac interferometer accompanies connecting an asymmetric-Mach-Zehnder structure (AMZS) that consists of a long and a short path is employed to modulate the four quantum states. The PM-2 in the Sagnac interferometer allows the users to modulate the interference of the Sagnac interferometer to switch the path. The constructive and destructive interference pipe the pulse into long and short paths respectively, where the long path case denotes the late bin occupation $\ket{0}$ and the short path case denotes the early bin occupation $\ket{1}$. A medium interference would split the pulse into both of the paths to prepare a superposition of $\ket{0}$ and $\ket{1}$ so that the test basis is selected in this case. In the AMZS, a phase modulator PM-3 is inserted into the short path to modulate the relative phase of the two paths. The code of the test basis is defined as the relative phase of the two paths that $0$ and $\pi$ denote the $\ket{\varphi_2}$ and $\ket{\varphi_3}$ respectively. The PM-2 and the PM-3 are also driven by the homemade 2-bit DAC to randomly modulate the four quantum states where the digital random numbers also come from the FPGA-based driver board.

Finally, the pulses are attenuated to a single-photon level by an electric variable optical attenuator (EVOA) and split to two parts, one of which is measured by a local single-photon detector (SPD) (SPD-300, Qasky), and the results are recorded by the FPGA-based driver board to monitor the intensity of the four decoy-states. According to the statistical result of the local SPD, users can adjust the EVOA and compensate for the DC drift of the IM-2 to keep the stability of the decoy states. The other part is sent to Charlie through a 25 km standard fiber for the BSM.

For the measurement unit Charlie, the two pulses interfere in the beam splitter (BS) and are detected by SPD-L and SPD-S (SPD-300, Qasky) that work in the gated mode \cite{qian2019robust}. The BS and the two SPDs constitute a BSM device for projecting quantum states to the bell state $\ket{\psi^-}$ \cite{lo2012measurement}. The two SPDs whose gate width, average efficiency, and dark count rate are 1 ns, $20.9\%$, and $3\times 10^{-6}$ respectively, are triggered by 25 MHz gate signals, and the gate signals for SPD-L and SPD-S are aligned with late bin occupation $\ket{0}$ and early bin occupation $\ket{1}$, respectively. The two electronic polarization controllers (EPCs) are employed to compensate for the polarization drift \cite{wang2015phase,ding2017polarization}. 

To demonstrate the imbalance tolerance, we experimentally demonstrate our protocol in three different imbalances as listed in Tab.\ref{Tab_experimental_imbalances} and calculate the secret key rate by the analysis methods for pure-state case and mixed-state case respectively. The system is continuously run to collect $5\times10^{11}$ pulse pairs for the case of each imbalance. The improved four-intensity decoy-state method against the effect of statistical fluctuation is proposed and employed for data processing (the details are in the Supplemental Material). With the improved decoy-state method, we successfully generate secret keys with the SKR of $1.10\times10^{-6}$, $7.37\times10^{-7}$, and $5.87\times10^{-7}$ by the "pure-state" analysis method and SKR of $8.56\times10^{-7}$, $6.39\times10^{-7}$, $3.87\times10^{-7}$ by the "mixed-state" analysis method. The maximum Z basis misalignment $\upsilon^U$ for the "mixed-state" analysis is set to $1\%$ (We note that we process the data by two different analysis methods to generate the SKR for the "pure-state" cases and the "mixed-state" cases ). 

\begin{figure}[htbp]
	\includegraphics[width=8cm]{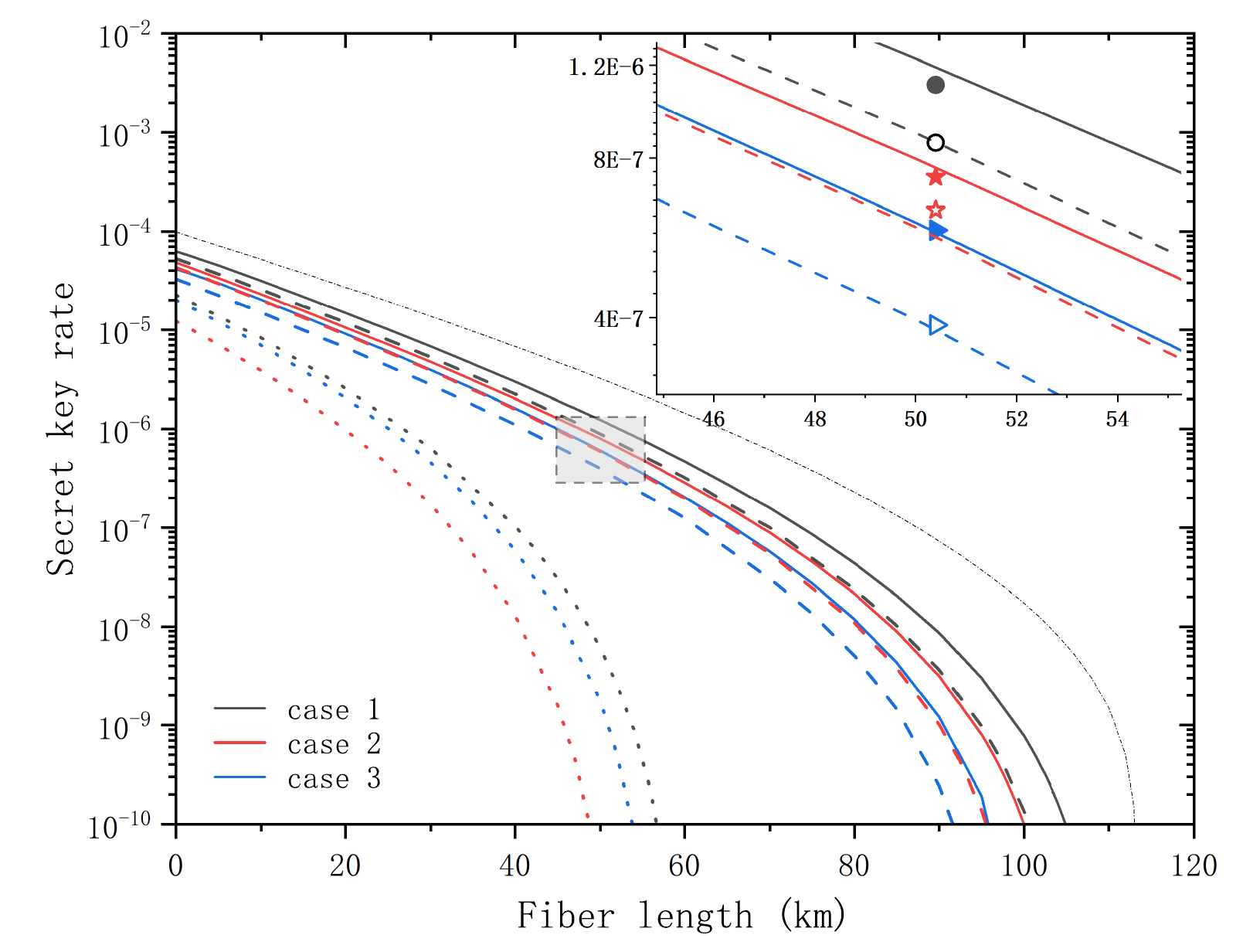}
	\caption{\label{Fig_experimental_results} The main figure illustrates non-asymptotic SKRs as the function of fiber length. The black, red, and blue denote the three different imbalances as listed in Tab.\ref{Tab_experimental_imbalances} respectively. The solid lines and dash lines denote our protocol with the pure-state scenario and the mixed-state scenario. The dot lines denote the state-of-the-art uncharacterized qubits MDIQKD \cite{hwang2017improved} and the thin dash-dot line denotes the original MDIQKD without any imbalances. The simulation parameters are $P_d = 3\times10^{-6}$, $P_\eta = 20\%$, the data size $N = 5\times10^{11}$ ,the background error rate is $0.9\%$, the failure probability $\epsilon$ in using chernoff bound \cite{zhang2017improved} is fixed to $5.73\times10^{-7}$, the fiber loss is 0.2 dB/km, the $\upsilon^L$ and $\upsilon^U$ for the "mixed-state" analysis are $0$ and $1\%$ respectively. The enlarged subfigure shows our experiment results. The solid geometries and hollow geometries denote the results for the pure-state scenario and the mixed-state scenario respectively.
 }
\end{figure}

Figure \ref{Fig_experimental_results} illustrates the simulation of the non-asymptotic SKRs (only including the statistical fluctuation) and shows our experiment results in the enlarged subfigure. The experiment results indicate that the protocol maintains a good performance against the unbalanced basis in the non-asymptotic cases. In non-asymptotic cases, we can find that our protocol performance is worse than the original MDIQKD and the SKR decrease with the imbalance $\beta$ and $\beta'$. The main reason is that our protocol estimates more quantities so that the impacts of the statistical fluctuations are more serious. A feasible solution for this problem is improving the decoy-state method \cite{yu2015statistical,zhou2016making,jiang2021higher,lu2020efficient}. Considering that our decoy-state is primitive, the protocol performance could be further improved in following works.

\begin{table}[htbp]
\caption{\label{Tab_experimental_imbalances}
Three different imbalances in our experiment }
\begin{ruledtabular}
\begin{tabular}{ccc}
Imbalances & Alice's imbalance $\beta$ & Bob's imbalance $\beta'$ \\
\hline
\textit{case 1} & 0 & 0   \\
\textit{case 2} & 0 & $10^{\circ}$ \\
\textit{case 3} & $10^{\circ}$ & $10^{\circ}$  \\
\end{tabular}
\end{ruledtabular}
\end{table}

\begin{widetext}

\section{Security proof}

\subsection{Pure-state scenario}
Alice and Bob prepare $N$ pairs of entangled states that
\begin{equation}
\begin{aligned}
\label{initial_state}
&\ket{\phi}_{A} = \sqrt{\frac{P_C}{2}}(\ket{0}_A\ket{\varphi_0} + \ket{1}_A\ket{\varphi_1}) + \sqrt{\frac{P_T}{2}}(\ket{2}_A\ket{\varphi_2} + \ket{3}_A\ket{\varphi_3}) , \\
&\ket{\phi}_{B} =  \sqrt{\frac{P'_C}{2}}(\ket{0}_B\ket{\varphi'_0} + \ket{1}_B\ket{\varphi'_1}) +\sqrt{\frac{P'_T}{2}}(\ket{2}_B\ket{\varphi'_2} + \ket{3}_B\ket{\varphi'_3}), \\
&P_C + P_T =1,\ \  P'_C + P'_T =1,
\end{aligned}
\end{equation} 
where the state $\ket{n}_A$ and $\ket{m}_B$ denote the ancillas of Alice and Bob respectively. $P_C$ and $P_T$ ($P'_C$ and $P'_T$) are probabilities for selecting code mode and test mode respectively. $n(m) = 0,1$ denotes the code basis, and $n(m) = 2,3$ denotes the test basis. The $\ket{\varphi_n}$ and $\ket{\varphi'_m}$ are pure states in a two-dimensional Hilbert space (qubit) to send to Charlie for BSM and they satisfy
\begin{equation}
\begin{aligned}
\label{state relations}
&\ket{\varphi_0} = \ket{\varphi'_0} = \ket{0},\ \ket{\varphi_1} = \ket{\varphi'_1} = \ket{1}, \\ 
&\ket{\varphi_2} = c_0 \ket{0} + c_1 \ket{1},\  \ket{\varphi_3} = c_0 \ket{0} - c_1 \ket{1},\\
&\ket{\varphi'_2} = c'_0 e^{i\theta} \ket{0} + c'_1 \ket{1},\ \ket{\varphi'_3} = c'_0 \ket{0} - c'_1 e^{i\theta} \ket{1},\\
\end{aligned}
\end{equation}	
The measurement unit Charlie can be anyone, so we consider the worst case that he is, in fact, the eavesdropper Eve. Following the similar argument used in Ref. \cite{shor2000simple}, we design an entanglement distillation protocol that is equivalent to the MDIQKD protocol \cite{lo2012measurement}. The initial states $\ket{\varphi_n}$, $\ket{\varphi'_m}$ and the Eve's ancillas $\ket{e}_E$ are separable. So Eve's collective attack can be defined as 
\begin{equation}
\begin{aligned}
\label{collective_attack_Z}
&U_{Eve}\ket{\varphi_n}\ket{\varphi'_m}\ket{e}_E\ket{0}_M = \sqrt{q_{nm}^0}\ket{\Gamma_{nm}^0 }_E\ket{0}_M + \sqrt{q_{nm}^1}\ket{\Gamma_{nm}^1 }_E\ket{1}_M,
\end{aligned}
\end{equation}
where the $\ket{0}_M$ and $\ket{1}_M$ is publicly announced message that $0$ is failure and $1$ is success; the $\ket{\Gamma_{nm}^{0(1)} }_E$ is the corresponding normalized Eve’s arbitrary quantum states for $\ket{\varphi_n}\ket{\varphi'_m}\ket{e}_E$ and the failure (success) event; the $q_{nm}^0$ and $q_{nm}^1$ are the corresponding probabilities of the $\ket{\Gamma_{nm}^{0} }_E$ and $\ket{\Gamma_{nm}^{1} }_E$ respectively. Considering that the failure events have been discarded, we only need to consider the terms $\sqrt{q_{nm}^1}\ket{\Gamma_{nm}^1 }_E\ket{1}_M$. The $\ket{1}_M$ and superscript $1$ could be omitted for simplification. Combining Eq.(\ref{initial_state}) and (\ref{collective_attack_Z}), the Alice, Bob and Eve's post-selected state of the code basis is 
\begin{equation}
\begin{aligned}
\label{post_selected_state}
\ket{\Phi}_{ABE}^Z = \frac{\sum_{n=0}^1\sum_{m=0}^1\sqrt{q_{nm}} \ket{\Gamma_{nm}}_E \ket{n}_A\ket{m}_B }
{ \sqrt{q_{00} + q_{01} + q_{10} + q_{11} } },
\end{aligned}
\end{equation}
where the $\ket{\Gamma_{nm}}_E$ can be described as an orthogonal decomposition in a set
of orthonormal states $\ket{v}$ that $\ket{\Gamma_{nm}}_E = \sum_v \gamma_{nm}^v\ket{v}$.
From normalization, $\sum_v |\gamma_{nm}^v|^2 = 1$.
The density matrix of Alice and Bob is 
\begin{equation}
\begin{aligned}
\label{density_matrix}
\rho = tr_E(\ket{\Phi}_{ABE}^Z\bra{\Phi}) =\frac{\sum_v\mathcal{P}(\sum_{n=0}^1\sum_{m=0}^1 \sqrt{q_{nm}}\gamma_{nm}\ket{n}_A\ket{m}_B )}{ q_{00} + q_{01} + q_{10} + q_{11} }
\end{aligned}
\end{equation}
where $tr_E$ denote the trace and $\mathcal{P}(\ket{k}) = \ket{k}\bra{k}$. 

The aim of the entanglement distillation protocol is to find a set of Bell states and project Alice's and Bob's ancillas to the $\ket{\psi^{-}}$. We define the Bell states as
\begin{equation}
\begin{aligned}
\label{density_matrices}
& \ket{\phi^{+}} = (\ket{0}_A \ket{0}_B + e^{i\theta} \ket{1}_A \ket{1}_B)/\sqrt{2},\\
& \ket{\phi^{-}} = (\ket{0}_A \ket{0}_B - e^{i\theta} \ket{1}_A \ket{1}_B)/\sqrt{2},\\
& \ket{\psi^{+}} = (\ket{0}_A \ket{1}_B + e^{-i\theta} \ket{1}_A \ket{0}_B)/\sqrt{2},\\
& \ket{\psi^{-}} = (\ket{0}_A \ket{1}_B - e^{-i\theta} \ket{1}_A \ket{0}_B)/\sqrt{2},\\
\end{aligned}
\end{equation}
thus the bit error rate $e_b$ and phase error rate $e_p$ of the code basis are
\begin{equation}
\begin{aligned}
\label{bit_error_rate}
e_b &= \bra{\phi^-}\rho\ket{\phi^-} + \bra{\phi^+}\rho\ket{\phi^+}\\
&= \frac{q_{00} + q_{11} }{q_{00} + q_{01} + q_{10} + q_{11}},
\end{aligned}
\end{equation}
and 
\begin{equation}
\begin{aligned}
\label{bit_error_rate}
e_p &= \bra{\psi^+}\rho\ket{\psi^+} + \bra{\phi^+}\rho\ket{\phi^+}\\
&= \frac{1}{2} + \frac{\sqrt{q_{00}q_{11} }\Re \left[ e^{i\theta}\bra{\Gamma_{00}}\ket{\Gamma_{11}} \right] + \sqrt{q_{01}q_{01} }\Re \left[ e^{-i\theta}\bra{\Gamma_{01}}\ket{\Gamma_{10}} \right]  }{q_{00} + q_{01} + q_{10} + q_{11}}.
\end{aligned}
\end{equation}
respectively.

The key idea for estimating information leakage is using the observable values to describe the $e_p$. Eq.(\ref{collective_attack_Z}) and Eq.(\ref{state relations}) give constraints that
\begin{equation}
\begin{aligned}
\label{collective_attack_X1}
&\sqrt{q_{2m}}\ket{\Gamma_{2m}} = \sqrt{q_{0m}}c_0\ket{\Gamma_{0m}} + \sqrt{q_{1m}}c_1\ket{\Gamma_{1m}},\\
&\sqrt{q_{3m}}\ket{\Gamma_{3m}} = \sqrt{q_{0m}}c_0\ket{\Gamma_{0m}} - \sqrt{q_{1m}}c_1\ket{\Gamma_{1m}},\\
&\sqrt{q_{n2}}\ket{\Gamma_{n2}} = \sqrt{q_{n0}}c'_0\ket{\Gamma_{n0}} + \sqrt{q_{n1}}c'_1 e^{i\theta}\ket{\Gamma_{n1}},\\
&\sqrt{q_{n3}}\ket{\Gamma_{n3}} = \sqrt{q_{n0}}c'_0\ket{\Gamma_{n0}} - \sqrt{q_{n1}}c'_1 e^{i\theta}\ket{\Gamma_{n1}},
\end{aligned}
\end{equation}
for $n,m\in\{0,1\}$, and 
\begin{equation}
\begin{aligned}
\label{collective_attack_X2}
&\sqrt{q_{22}}\ket{\Gamma_{22}} = \sum_{n,m\in\{0,1\}}\sqrt{q_{nm}}c_n c'_m e^{i\theta_m}\ket{\Gamma_{nm}},\\
&\sqrt{q_{23}}\ket{\Gamma_{23}} = \sum_{n,m\in\{0,1\}}(-1)^m\sqrt{q_{nm}}c_n c'_m e^{i\theta_m}\ket{\Gamma_{nm}},\\
&\sqrt{q_{32}}\ket{\Gamma_{32}} = \sum_{n,m\in\{0,1\}}(-1)^n\sqrt{q_{nm}}c_n c'_m e^{i\theta_m}\ket{\Gamma_{nm}},\\
&\sqrt{q_{33}}\ket{\Gamma_{33}} = \sum_{n,m\in\{0,1\}}(-1)^{n+m}\sqrt{q_{nm}}c_n c'_m e^{i\theta_m}\ket{\Gamma_{nm}},\\
\end{aligned}
\end{equation}
where $\theta_m = 0$ when $m=0$ and $\theta_m = \theta$ when $m=1$.

Eq.(\ref{collective_attack_X1}) indicates that 
\begin{equation}
\begin{aligned}
&q_{2m} = q_{0m}c_0^2 + q_{1m}c_1^2 + \sqrt{q_{0m}q_{1m}}c_0c_1\big(\bra{\Gamma_{0m}}\ket{\Gamma_{1m}} + \bra{\Gamma_{1m}}\ket{\Gamma_{0m}}\big),\\
&q_{3m} = q_{0m}c_0^2 + q_{1m}c_1^2 - \sqrt{q_{0m}q_{1m}}c_0c_1\big(\bra{\Gamma_{0m}}\ket{\Gamma_{1m}} + \bra{\Gamma_{1m}}\ket{\Gamma_{0m}}\big),\\
&q_{n2} = q_{n0}{c'_0}^2 + q_{n1}{c'_1}^2 + \sqrt{q_{n0}q_{n1}}c'_0 c'_1\big(e^{i\theta}\bra{\Gamma_{n0}}\ket{\Gamma_{n1}} + e^{-i\theta}\bra{\Gamma_{n1}}\ket{\Gamma_{n0}}\big),\\
&q_{n3} = q_{n0}{c'_0}^2 + q_{n1}{c'_1}^2 - \sqrt{q_{n0}q_{n1}}c'_0c'_1\big(e^{i\theta}\bra{\Gamma_{n0}}\ket{\Gamma_{n1}} + e^{-i\theta}\bra{\Gamma_{n1}}\ket{\Gamma_{n0}}\big),\\
\end{aligned}
\end{equation}
so that
\begin{equation}
\begin{aligned}
\label{c_relations}
&2q_{0m}{c_0}^2 + 2q_{1m}{c_1}^2 = q_{2m} + q_{3m},\\
&2q_{n0}{c'_0}^2 + 2q_{n1}{c'_1}^2 = q_{n2} + q_{n3}.
\end{aligned}
\end{equation}
The positive real coefficients $c_0$, $c_1$, $c'_0$, $c'_1$ can be accurately calculated.
Besides, Eq.(\ref{collective_attack_X2}) indicates that
\begin{equation}
\begin{aligned}
\label{q_nm of test basis}
&q_{22} = \sum_{n,m,k,l\in\{0,1\}}^{nm\neq kl} \sqrt{q_{nm}q_{kl}}c_n c'_m c_k c'_l \bra{\Gamma_{kl}}\ket{\Gamma_{nm}}e^{i(\theta_m-\theta_l)},\\
&q_{23} = \sum_{n,m,k,l\in\{0,1\}}^{nm\neq kl} (-1)^{m+l}\sqrt{q_{nm}q_{kl}}c_n c'_m c_k c'_l \bra{\Gamma_{kl}}\ket{\Gamma_{nm}}e^{i(\theta_m-\theta_l)},\\
&q_{32} = \sum_{n,m,k,l\in\{0,1\}}^{nm\neq kl} (-1)^{n+k}\sqrt{q_{nm}q_{kl}}c_n c'_m c_k c'_l \bra{\Gamma_{kl}}\ket{\Gamma_{nm}}e^{i(\theta_m-\theta_l)},\\
&q_{33} = \sum_{n,m,k,l\in\{0,1\}}^{nm\neq kl} (-1)^{n+m+k+l}\sqrt{q_{nm}q_{kl}}c_n c'_m c_k c'_l \bra{\Gamma_{kl}}\ket{\Gamma_{nm}}e^{i(\theta_m-\theta_l)},
\end{aligned}
\end{equation}
where $\theta_{m(l)} = 0$ when $m(l)=0$ and $\theta_{m(l)}=\theta$ when $m(l)=1$. Accoding to Eq.(\ref{q_nm of test basis}), we can find that
\begin{equation}
\begin{aligned}
\label{c_relations}
 - (q_{23} + q_{32}) + (q_{22} +q_{33} ) = 8 c_0c'_0c_1c'_1 \left[\sqrt{q_{00}q_{11}}\Re(e^{i\theta}\bra{\Gamma_{11}}\ket{\Gamma_{00}}) + \sqrt{q_{10}q_{01}} \Re(e^{-i\theta}\bra{\Gamma_{10}}\ket{\Gamma_{01}})\right],
\end{aligned}
\end{equation}
so that
\begin{equation}
\begin{aligned}
\label{phase_error}
e_p = \frac{1}{2} - \frac{(q_{23}+q_{23}) - (q_{22}+q_{33})}{8(q_{00} +q_{01} +q_{10} +q_{11})c_0 c'_0 c_1 c'_1},
\end{aligned}
\end{equation}

Indeed, Charlie may not project his received pulse pair to the $\ket{\psi^{-}} = (\ket{0} \ket{1} - e^{-i\theta} \ket{1} \ket{0})/\sqrt{2}$ because he is unreliable. However, according to the property of the MDI-QKD, the protocol security has nothing to do with Charlie's measurement and announcement. So our protocol is secure even if the phase reference frame between Alice and Bob is not well-aligned. Besides, if Charlie is honest, he can compensate the reference-frame misalignment $\theta$ to maximize the secret key rate. \\ \\

\subsection{modulation fluctuation in the test basis}

In practical system, the unbalanced-basis-misalignments $\beta$ and $\beta'$ (we will abbreviate it to "imbalance" for simplicity) are usually not a fixed values but satisify the random distribution with probability density function $P(\beta)$ due to the modulation fluctuations. The key idea for proving the security against this case is equating the source side modulation fluctuations to Eve's operations in quantum channel. Here we introduce the method to equating the mixed state to uncharacterized $\ket{\varphi_2}$ and $\ket{\varphi_3}$ that satisfy Eq.(\ref{state relations}). We define
\begin{equation}
\begin{aligned}
&\mathcal{P}\left(\cos(45^\circ + \beta)\ket{0} + \sin(45^\circ + \beta)e^{i\theta}\ket{1}\right) = \frac{1}{2}(\mathbb{I} + S_X(\beta)\sigma^\theta_X + S_Z(\beta)\sigma_Z ),\\
&\mathcal{P}\left(\cos(45^\circ + \beta)\ket{0} - \sin(45^\circ + \beta)e^{i\theta}\ket{1}\right) = \frac{1}{2}(\mathbb{I} - S_X(\beta)\sigma^\theta_X + S_Z(\beta)\sigma_Z ),\\
\end{aligned}
\end{equation}
where $\mathcal{P}(\ket{x}) = \ket{x}\bra{x}$, $\mathbb{I}$ is the identity matrix, and
\begin{equation}
\begin{aligned}
&\sigma_Z = \ket{0}\bra{0} - \ket{1}\bra{1},\\
&\sigma^\theta_X = \mathcal{P}(\frac{\ket{0} + e^{i\theta}\ket{1}}{\sqrt{2}}) - \mathcal{P}(\frac{\ket{0} - e^{i\theta}\ket{1}}{\sqrt{2}}),
\end{aligned}
\end{equation}
are Pauli matrices.

The test basis can be described as mixed states
\begin{equation}
\begin{aligned}
& \rho_2 = \int^{\overline{\beta}}_{\underline{\beta}}  \frac{1}{2}P(\beta)(\mathbb{I} + S_X(\beta)\sigma^\theta_X + S_Z(\beta)\sigma_Z ) \rm{d}\beta
 = \frac{1}{2}(\mathbb{I} + \hat{S}_X\sigma^\theta_X + \hat{S}_Z\sigma_Z )
\end{aligned}
\end{equation}
and
\begin{equation}
\begin{aligned}
 &\rho_3 = \int^{\overline{\beta}}_{\underline{\beta}}  \frac{1}{2}P(\beta)(\mathbb{I} - S_X(\beta)\sigma^\theta_X + S_Z(\beta)\sigma_Z ) \rm{d}\beta
 = \frac{1}{2}(\mathbb{I} - \hat{S}_X\sigma^\theta_X + \hat{S}_Z\sigma_Z )
 \end{aligned}
\end{equation}
where $\int^{\overline{\beta}}_{\underline{\beta}} P(\beta) \mathrm{d}\beta = 1$, $\hat{S}_X = \int^{\overline{\beta}}_{\underline{\beta}} S_X(\beta) P(\beta) \mathrm{d}\beta$, $\hat{S}_Z = \int^{\overline{\beta}}_{\underline{\beta}} S_Z(\beta) P(\beta) \mathrm{d}\beta$.

We define $\ket{\varphi_2} = c_0\ket{0} + c_1e^{i\theta}\ket{1}$ and $\ket{\varphi_3} = c_0\ket{0} - c_1e^{i\theta}\ket{1}$ are two uncharacterized states satisify
\begin{equation}
\begin{aligned}
 &\ket{\varphi_2}\bra{\varphi_2} = \frac{1}{2}\left( \mathbb{I} + \sqrt{1 - {\hat{S}_Z}^2}\sigma^\theta_X + \hat{S}_Z \sigma_Z  \right),\\
 &\ket{\varphi_3}\bra{\varphi_3} = \frac{1}{2}\left( \mathbb{I} - \sqrt{1 - {\hat{S}_Z}^2}\sigma^\theta_X + \hat{S}_Z \sigma_Z  \right).
 \end{aligned}
\end{equation}

Define Eve's operation as
\begin{equation}
\varepsilon(\rho) = \lambda_0 \mathbb{I}\rho\mathbb{I} + \lambda_1 \sigma_Z \rho \sigma_Z, 
\end{equation}
which satisfies
\begin{equation}
\begin{aligned}
\left \{
\begin{array}{lc}
\lambda_0 + \lambda_1 = 1,\\
(\lambda_0 - \lambda_1) \sqrt{1 - {\hat{S}_Z}^2} = \hat{S}_X.
\end{array}
\right.
\end{aligned}
\end{equation}
The $\varepsilon$ can be intuitively regarded as Eve doing nothing with the probability $\lambda_0$ and performing the $\sigma_Z$-operation with the probability $\lambda_1$. We can find that the $\varepsilon$ satisfies
\begin{equation}
\begin{aligned}
& \ket{0}\bra{0} = \varepsilon(\ket{0}\bra{0}), \ \ \ket{1}\bra{1} = \varepsilon(\ket{1}\bra{1}),\\
& \rho_2 = \varepsilon \left(\ket{\varphi_2}\bra{\varphi_2}\right), \ \   \rho_3 = \varepsilon \left(\ket{\varphi_3}\bra{\varphi_3}\right).
 \end{aligned}
\end{equation}

So the mixed states $\rho_2$ and $\rho_3$ can be regarded as uncharacterized pure states $\ket{\varphi_2}$ and $\ket{\varphi_3}$ with Eve's operation $\varepsilon$ in the quantum channel, and the $\ket{\varphi_2}$ and $\ket{\varphi_3}$ satisfy Eq.(\ref{state relations}). Noting that the operations in the quantum channel have nothing to do with the security, we can claim that the security in this scenario equals to the pure-state scenario.

\subsection{modulation misalignment and fluctuation in the code basis} 

We note that Z basis of the time-bin phase coding systems is usually well-aligned so that its modulation misalignment is ignorable and the above pure-state method is enough for most of the scenarios. However, if the Z basis is lose control due to the modulation error and fluctuation, we should consider an improved post-processing. 

Our key idea is to bound the pure-state yields by the practically observable mixed-state yields. Noting that the preparing the mixed states in code basis is equivalent to the case that Alice (Bob) prepare pure states normally but randomly adds some noise to her (his) Z-basis raw key bits. In other words, Alice (Bob) randomly flips some Z-basis raw key bits and forgets which bits are flipped. Compared with the original pure-state protocol in which there is no the random flips, Eve obviously cannot get any additional information. So the information leakage upper bound for the pure-state protocol still holds in the mixed-state case.

In the scenarios that the Z basis also has misalignments, Alice actually randomly prepares $\cos( \beta_0 )\ket{0} + e^{i\theta}\sin( \beta_0 )\ket{1}$ or $\cos( \beta_0 )\ket{0} - e^{i\theta}\sin( \beta_0 )\ket{1}$ when she wants to prepare $\ket{0}$ and actually prepares $\sin( \beta_1 )\ket{0} + e^{i\theta}\cos( \beta_1 )\ket{1}$ or $\sin( \beta_1 )\ket{0} - e^{i\theta}\cos( \beta_1 )\ket{1}$ when she wants to prepare $\ket{1}$. The randomness of above "$+$" and "$-$" derive from Alice randomly modulate $0$ or $\pi$ relative phase between the two time bins when selecting the Z basis. Besides, because of the modulation fluctuations, the misalignment $\beta_0$ and $\beta_1$ are not fixed values but satisify the random distributions with probability density functions $P_0(\beta_0)$ and $P_1(\beta_1)$ respectively. In other words, Alice actually prepares mixed states 
\begin{equation}
\begin{aligned}
& \rho_0 = \int^{\overline{\beta}_0}_{\underline{\beta}_0} P_0(\beta_0) \frac{1}{2}\Big( \mathcal{P}\left(\cos( \beta_0)\ket{0} + \sin( {\beta}_0)e^{i\theta}\ket{1}\right) + \mathcal{P}\left(\cos( \beta_0)\ket{0} - \sin( {\beta}_0)e^{i\theta}\ket{1}\right)\Big)  \mathrm{d}{\beta}_0 \\
& = \int^{\overline{\beta}_0}_{\underline{\beta}_0} P_0(\beta_0) \left( \cos^2( \beta_0) \ket{0}\bra{0} + \sin^2( {\beta}_0)\ket{1}\bra{1} \right) \mathrm{d}{\beta}_0 \\ \\
&  =(1 - \xi)\ket{0}\bra{0} + \xi \ket{1}\bra{1},
\end{aligned}
\end{equation}
when she wants to prepare $\ket{0}$, and 
\begin{equation}
\begin{aligned}
& \rho_1 = \int^{\overline{\beta}_1}_{\underline{\beta}_1} P_1(\beta_1) \frac{1}{2}\Big( \mathcal{P}\left(\sin( \beta_1)\ket{0} + \cos( {\beta}_1)e^{i\theta}\ket{1}\right) + \mathcal{P}\left(\sin( \beta_1)\ket{0} - \cos( {\beta}_1)e^{i\theta}\ket{1}\right)\Big)  \mathrm{d}{\beta}_1 \\
& = \int^{\overline{\beta}_1}_{\underline{\beta}_1} P_1(\beta_1) \left( \sin^2( \beta_1) \ket{0}\bra{0} + \cos^2( {\beta}_1)\ket{1}\bra{1} \right) \mathrm{d}{\beta}_1 \\ \\
& = \zeta\ket{0}\bra{0} + (1-\zeta) \ket{1}\bra{1},
\end{aligned}
\end{equation}
when she wants to prepare $\ket{1}$. The misalignment error $\xi$ and $\zeta$ are defined as
\begin{equation}
\begin{aligned}
\xi = \int^{\overline{\beta}_0}_{\underline{\beta}_0} P_0(\beta_0) \sin^2(\beta_0) \mathrm{d}{\beta}_0, 
\end{aligned}
\end{equation}
and 
\begin{equation}
\begin{aligned}
\zeta = \int^{\overline{\beta}_1}_{\underline{\beta}_1} P_1(\beta_1) \sin^2(\beta_1) \mathrm{d}{\beta}_1, 
\end{aligned}
\end{equation}
respectively. Similarly, when Bob wants to prepare $\ket{0}$ and $\ket{1}$, he actually prepares mixed states
\begin{equation}
\rho'_0 = (1 - \xi')\ket{0}\bra{0} + \xi' \ket{1}\bra{1},
\end{equation}
and 
\begin{equation}
\rho'_1 = \zeta' \ket{0}\bra{0} + (1-\zeta') \ket{1}\bra{1},
\end{equation}
respectively, where $\xi'$ and $\zeta'$ are misalignment errors of the Z basis. When both of the users select $\ket{0}$, they actually prepare
\begin{equation}
\begin{aligned}
\label{joint density matrix1}
&\rho_0\otimes\rho'_0 =\\ & (1-\xi)(1-\xi')\ket{0}\bra{0}\otimes \ket{0}\bra{0} + (1-\xi)\xi' \ket{0}\bra{0} \otimes \ket{1}\bra{1}  +\\ & \xi(1-\xi') \ket{1}\bra{1} \otimes \ket{0}\bra{0}+  \xi\xi'\ket{1}\bra{1} \otimes \ket{1}\bra{1}
\end{aligned}
\end{equation}
Eq.(\ref{joint density matrix1}) indicates that
\begin{equation}
y_{00} = (1-\xi)(1-\xi')q_{00} + (1-\xi)\xi'q_{01} + \xi(1-\xi')q_{10} + \xi\xi'q_{11}.
\end{equation}
Similarly, we have
\begin{equation}
\begin{aligned}
&\rho_0\otimes\rho'_1 =\\ & (1-\xi)\zeta'\ket{0}\bra{0}\otimes \ket{0}\bra{0} + (1-\xi)(1-\zeta') \ket{0}\bra{0} \otimes \ket{1}\bra{1}  +\\ & \xi\zeta' \ket{1}\bra{1} \otimes \ket{0}\bra{0}+  \xi(1-\zeta') \ket{1}\bra{1} \otimes \ket{1}\bra{1},
\end{aligned}
\end{equation}
\begin{equation}
\begin{aligned}
&\rho_1\otimes\rho'_0 =\\ & \zeta(1-\xi')\ket{0}\bra{0}\otimes \ket{0}\bra{0} + \zeta\xi' \ket{0}\bra{0} \otimes \ket{1}\bra{1}  +\\ & (1-\zeta)(1-\xi') \ket{1}\bra{1} \otimes \ket{0}\bra{0}+  (1-\zeta)\xi' \ket{1}\bra{1} \otimes \ket{1}\bra{1},
\end{aligned}
\end{equation}
and
\begin{equation}
\begin{aligned}
&\rho_1\otimes\rho'_1 =\\ & \zeta\zeta' \ket{0}\bra{0}\otimes \ket{0}\bra{0} + \zeta(1-\zeta') \ket{0}\bra{0} \otimes \ket{1}\bra{1}  +\\ & (1-\zeta)\zeta' \ket{1}\bra{1} \otimes \ket{0}\bra{0}+  (1-\zeta)(1-\zeta') \ket{1}\bra{1} \otimes \ket{1}\bra{1},
\end{aligned}
\end{equation}
which indicate that
\begin{equation}
\label{Eq Z basis with misalignment}
\begin{aligned}
\left( 
\begin{array}{cccc}
(1-\xi)(1-\xi') & (1-\xi)\xi' & \xi(1-\xi') & \xi\xi' \\
(1-\xi)\zeta' & (1-\xi)(1-\zeta') & \xi\zeta' & \xi(1-\zeta') \\
\zeta(1-\xi') & \zeta\xi' & (1-\zeta)(1-\xi') & (1-\zeta)\xi' \\
\zeta\zeta' & \zeta(1-\zeta') & (1-\zeta)\zeta' & (1-\zeta)(1-\zeta')\\
\end{array}
\right)
\left(
\begin{array}{c}
q_{00}\\
q_{01}\\
q_{10}\\
q_{11}\\
\end{array}
\right)
=
\left(
\begin{array}{c}
y_{00}\\
y_{01}\\
y_{10}\\
y_{11}\\
\end{array}
\right).
\end{aligned}
\end{equation}
By solving the above equation set, we can get
\begin{equation}
\begin{aligned}
\label{q_nm solutions}
&q_{00}=[(1-\zeta)(1-\zeta')y_{00}-(1-\zeta)\xi y_{01}-\xi (1-\zeta') y_{10} +\xi \xi' y_{11}]/[(1-\xi-\zeta)(1-\xi'-\zeta')],\\
&q_{01}=[-(1-\zeta)\zeta'y_{00}+(1-\zeta)(1-\xi') y_{01}+\xi \zeta' y_{10} -\xi (1-\xi') y_{11}]/[(1-\xi-\zeta)(1-\xi'-\zeta')],\\
&q_{10}=[-\zeta(1-\zeta')y_{00}+\zeta \xi' y_{01}+(1-\xi) (1-\zeta') y_{10} -(1-\xi) \xi' y_{11}]/[(1-\xi-\zeta)(1-\xi'-\zeta')],\\
&q_{11}=[\zeta\zeta'y_{00}-\zeta(1-\xi') y_{01}-(1-\xi)\zeta' y_{10} +(1-\xi) (1-\xi') y_{11}]/[(1-\xi-\zeta)(1-\xi'-\zeta')].\\
\end{aligned}
\end{equation}
Indeed, Alice and Bob do not know the specific value of the misalignments $\zeta$, $\zeta'$, $\xi$, and $\xi'$, but they can estimate the lower and upper bounds of these misalignments according to their calibration or monitor these misalignments by a local detector similar to the monitor "SPD-Moni" in our experiment. We define $\upsilon^L$ and $\upsilon^U$ as the estimated lower and upper bounds of misalignment $\upsilon$ where $\upsilon\in\{ \zeta, \zeta', \xi, \xi' \}$, the upper and lower bounds of the $q_{nm}$ can be estimated by considering the worst case of Eq.(\ref{q_nm solutions}) so that
\begin{equation}
\begin{aligned}
\label{q_nm bounds ZZ}
&q_{00}^L=\frac{(1-\zeta^U)(1-\zeta'^U)y_{00}-(1-\zeta^L)\xi^Uy_{01}-\xi^U(1-\zeta'^L)y_{10}+\xi^L\xi'^L y_{11}}{(1-\xi^L-\zeta^L)(1-\xi'^L-\zeta'^L)},\\
&q_{00}^U=\frac{(1-\zeta^L)(1-\zeta'^L)y_{00}-(1-\zeta^U)\xi^Ly_{01}-\xi^L(1-\zeta'^U)y_{10}+\xi^U\xi'^U y_{11}}{(1-\xi^U-\zeta^U)(1-\xi'^U-\zeta'^U)},\\
&q_{01}^L=\frac{-(1-\zeta^L)\zeta'^Uy_{00}+(1-\zeta^U)(1-\xi'^U)y_{01}+\xi^L\zeta'^Ly_{10}-\xi^U(1-\xi'^L)y_{11}}{(1-\xi^L-\zeta^L)(1-\xi'^L-\zeta'^L)},\\
&q_{01}^U=\frac{-(1-\zeta^U)\zeta'^Ly_{00}+(1-\zeta^L)(1-\xi'^L)y_{01}+\xi^U\zeta'^Uy_{10}-\xi^L(1-\xi'^U)y_{11}}{(1-\xi^U-\zeta^U)(1-\xi'^U-\zeta'^U)},\\
&q_{10}^L=\frac{-\zeta^U(1-\zeta'^L)y_{00}+\zeta^L\xi'^Ly_{01}+(1-\xi^U)(1-\zeta'^U)y_{10}-(1-\xi^L)\xi'^Uy_{11}}{(1-\xi^L-\zeta^L)(1-\xi'^L-\zeta'^L)},\\
&q_{10}^U=\frac{-\zeta^L(1-\zeta'^U)y_{00}+\zeta^U\xi'^Uy_{01}+(1-\xi^L)(1-\zeta'^L)y_{10}-(1-\xi^U)\xi'^Ly_{11}}{(1-\xi^U-\zeta^U)(1-\xi'^U-\zeta'^U)},\\
&q_{11}^L=\frac{\zeta^L\zeta'^Ly_{00}-\zeta^U(1-\xi'^L)y_{01}-(1-\xi^L)\zeta'^Uy_{10}+(1-\xi^U)(1-\xi'^U)y_{11}}{(1-\xi^L-\zeta^L)(1-\xi'^L-\zeta'^L)},\\
&q_{11}^U=\frac{\zeta^U\zeta'^Uy_{00}-\zeta^L(1-\xi'^U)y_{01}-(1-\xi^U)\zeta'^Ly_{10}+(1-\xi^L)(1-\xi'^L)y_{11}}{(1-\xi^U-\zeta^U)(1-\xi'^U-\zeta'^U)},\\
\end{aligned}
\end{equation}

The pure-state yields for the two users selecting different bases can also be estimated by solving equations.
We take the $q_{02}$ and $q_{12}$ as examples. We have proved that the fluctuation in the test basis can be equivalent to Eve's operations in the quantum channel so that we can always assume that the users prepare the uncharacterized pure-states $\ket{\varphi_2}$ and $\ket{\varphi_3}$. When Alice selects $\ket{0}$ and Bob selects $\ket{\varphi_2}$, they practically prepared mixed-state
\begin{equation}
\begin{aligned}
\label{q_nm solutions ZX basis}
\rho_0 \otimes \ket{\varphi_2}\bra{\varphi_2} 
= (1-\xi) \ket{0}\bra{0} \otimes \ket{\varphi_2}\bra{\varphi_2} 
+ \xi \ket{1}\bra{1} \otimes \ket{\varphi_2}\bra{\varphi_2}, 
\end{aligned}
\end{equation}
which indicates that
\begin{equation}
\begin{aligned}
y_{02} = (1-\xi) q_{02} + \xi q_{12}.
\end{aligned}
\end{equation}
Similarly, we have
\begin{equation}
\begin{aligned}
y_{12} = \zeta q_{02} + (1 - \zeta) q_{12},
\end{aligned}
\end{equation}
so that we can get an equation set that
\begin{equation}
\begin{aligned}
\left(
\begin{array}{cc}
1-\xi & \xi \\
\zeta & 1-\zeta \\
\end{array}
\right)
\left(
\begin{array}{c}
q_{02} \\
q_{12} \\
\end{array}
\right)
=
\left(
\begin{array}{c}
y_{02} \\
y_{12} \\
\end{array}
\right),
\end{aligned}
\end{equation}

The solution of the equation set is 
\begin{equation}
\begin{aligned}
&q_{02}=[(1-\zeta)y_{02}-\xi y_{12}]/(1-\xi-\zeta),\\
&q_{12}=[(1-\xi)y_{12}-\zeta y_{02}]/(1-\xi-\zeta),\\
\end{aligned}
\end{equation}
Considering the worst cases, the lower and upper bound of $q_{02}$ and $q_{12}$ are 
\begin{equation}
\begin{aligned}
&q_{02}^L=\frac{(1-\zeta^U)y_{02}-\xi^Uy_{12}}{1-\xi^U-\zeta^U},\ \
q_{02}^U=\frac{(1-\zeta^L)y_{02}-\xi^Ly_{12}}{1-\xi^L-\zeta^L},\\
&q_{12}^L=\frac{(1-\xi^U)y_{21}-\zeta^Uy_{20}}{1-\xi^U-\zeta^U},\ \
q_{12}^U=\frac{(1-\xi^L)y_{21}-\zeta^Ly_{20}}{1-\xi^L-\zeta^L},\\
\end{aligned}
\end{equation}
respectively. Similarly, we can conclude that 
\begin{equation}
\begin{aligned}
\label{q_nm bounds ZX}
&q_{0m}^L=\frac{(1-\zeta^U)y_{0m}-\xi^Uy_{1m}}{1-\xi^U-\zeta^U},\ \
q_{0m}^U=\frac{(1-\zeta^L)y_{0m}-\xi^Ly_{1m}}{1-\xi^L-\zeta^L},\\
&q_{1m}^L=\frac{(1-\xi^U)y_{m1}-\zeta^Uy_{m0}}{1-\xi^U-\zeta^U},\ \
q_{1m}^U=\frac{(1-\xi^L)y_{m1}-\zeta^Ly_{m0}}{1-\xi^L-\zeta^L},\\
&q_{n0}^L=\frac{(1-\zeta'^U)y_{n0}-\xi'^Uy_{n1}}{1-\xi'^U-\zeta'^U},\ \
q_{n0}^U=\frac{(1-\zeta'^L)y_{n0}-\xi'^Ly_{n1}}{1-\xi'^L-\zeta'^L},\\
&q_{n1}^L=\frac{(1-\xi'^U)y_{n1}-\zeta'^Uy_{n0}}{1-\xi'^U-\zeta'^U},\ \
q_{n1}^U=\frac{(1-\xi'^L)y_{n1}-\zeta'^Ly_{n0}}{1-\xi'^L-\zeta'^L},\\
\end{aligned}
\end{equation}
where $n,m\in\{2,3\}$. 

Besides, when both of the users select the test basis, we can conclude that 
$q_{nm} = y_{nm} (n,m\in\{2,3\})$ 
because the states in the test basis can be regarded as pure-states as we have proved. As far as we have bounded the $q_{nm}$, namely, the pure-state yields, the phase error $e_p$ can be estimated similar to the above pure-state cases.


\section{Four-intensity decoy state method}

\subsection{Pure-state scenario }

The above proofs are based on the ideal single-photon source that is still not practically usable. 
The decoy-state method \cite{hwang2003quantum,wang2005beating,lo2005decoy,yu2013three,yu2015statistical,zhou2016making} accompanying with coherent sources is widely employed as a substitute for the ideal single-photon source. Here we propose an efficient four-intensity decoy state method to solve the non-asymptotic cases. In our method, Alice (Bob) randomly prepares phase randomized weak coherent states with intensity $l$ ($r$) from a pre-decided set $\{\mu, \nu, \omega, o\}$ ($\{\mu', \nu', \omega', o\}$) where $\mu$ ($\mu'$) is defined as the signal state and the others are decoy states, and the decoy states meet $\nu > \omega$ and $o = 0$. If the signal state is selected, Alice (Bob) only selects the code basis, and if other intensities are selected, they randomly select the two bases just like the single-photon protocol. The $N_{nm}^{lr}$, $n_{nm}^{lr}$, and $Q_{nm}^{lr} = n_{nm}^{lr}/N_{nm}^{lr}$ denote the number of pulses pairs, the number of successful event, and the gain respectively. The superscript $lr$ denotes Alice and Bob selects intensity $l$ and $r$ respectively, and the subscript $nm$ denotes Alice and Bob prepare $\ket{\varphi_n}$ and $\ket{\varphi'_m}$ respectively. As a countermeasure of statistical fluctuation, we define some joint events whose gains are 
\begin{equation}
\begin{aligned}
&Q_{T_s}^{lr}=\frac{n_{22}^{lr}+n_{33}^{lr}}{N_{22}^{lr}+N_{33}^{lr}},\ \ 
Q_{T_d}^{lr}=\frac{n_{23}^{lr}+n_{32}^{lr}}{N_{23}^{lr}+N_{32}^{lr}},\\
&Q_{T0}^{lr}=\frac{n_{20}^{lr}+n_{30}^{lr}}{N_{20}^{lr}+N_{30}^{lr}}, \ \
Q_{T1}^{lr}=\frac{n_{21}^{lr}+n_{31}^{lr}}{N_{21}^{lr}+N_{31}^{lr}},\\
&Q_{0T}^{lr}=\frac{n_{02}^{lr}+n_{03}^{lr}}{N_{02}^{lr}+N_{03}^{lr}},\ \ 
Q_{1T}^{lr}=\frac{n_{12}^{lr}+n_{13}^{lr}}{N_{12}^{lr}+N_{13}^{lr}},\\
&Q_{C}^{lr}=\frac{n_{00}^{lr}+n_{01}^{lr}+n_{10}^{lr}+n_{11}^{lr}}{N_{00}^{lr}+N_{01}^{lr}+N_{10}^{lr}+N_{11}^{lr}},\\
\end{aligned}
\end{equation}
and the corresponding yields of single-photon pairs are
\begin{equation}
\begin{aligned}
&q_{T_s}=\frac{q_{22}+q_{33}}{2},
q_{T_d}=\frac{q_{23}+q_{32}}{2},
q_{T0}=\frac{q_{20}+q_{30}}{2},
q_{T1}=\frac{q_{21}+q_{31}}{2},\\
&q_{0T}=\frac{q_{02}+q_{03}}{2},
q_{1T}=\frac{q_{12}+q_{13}}{2},
q_{C}=\frac{q_{00}+q_{01}+q_{10}+q_{11}}{4}.
\end{aligned}
\end{equation}
The upper and lower bounds of the yields $q_\alpha$ for $\alpha\in\{00,01,10,11,T_s,T_d,T0,T1,0T,1T,C\}$ are estimated by \cite{yu2015statistical,zhou2016making,lu2020efficient}
\begin{equation}\label{q11}
\begin{aligned}
&q_{\alpha} \geq q_{\alpha}^L=\left( S_+ -  S_-\right) /\big[a_1^\omega a_1^\nu (b_1^\omega b_2^\nu - b_1^\nu b_2^\omega )\big],\\
&q_{\alpha} \leq q_{\alpha}^U=\left( S'_+ -  S'_- \right)/a_1^\omega b_1^\omega,
\end{aligned}
\end{equation}
where $a_k^l=l^ke^{-l}/l!$($b_k^r=r^ke^{-r}/r!$) is photon number distribution of coherent state with intensity $l(r)$, and
\begin{equation}
\begin{aligned}
\label{Eq_S}
&  S_+ ={a_1^\nu b_2^\nu}Q_\alpha^{\omega\omega} + {a_1^\omega b_2^\omega a_0^\nu}Q_\alpha^{o\nu} +{a_1^\omega b_2^\omega b_0^\nu}Q_\alpha^{\nu o} +{a_1^\nu b_2^\nu a_0^\omega b_0^\omega }Q_\alpha^{oo}  , \\
&  S_- ={a_1^\omega b_2^\omega}Q_\alpha^{\nu\nu} +{a_1^\nu b_2^\nu a_0^\omega}Q_\alpha^{o\omega} +{a_1^\nu b_2^\nu b_0^\omega}Q_\alpha^{\omega o} + a_1^\omega b_2^\omega a_0^\nu b_0^\nu Q_\alpha^{oo},\\
& S'_+ =  Q_\alpha^{\omega\omega} +{a_0^\omega b_0^\omega}Q_\alpha^{oo} ,\ \ 
 S'_- ={a_0^\omega}Q_\alpha^{o\omega} +{b_0^\omega}Q_\alpha^{\omega o} .
\end{aligned}
\end{equation}

The above Eq.(\ref{q11}) and Eq.(\ref{Eq_S}) are suitable for the asymptotic scenario that the data size $N\to\infty$. In non-asymptotic scenarios, the differences between observed values and expected values should be taken into consideration. A tight upper and lower bound of the yields can be estimated by optimizing ${n}_\alpha^{lr}$ in the linear programming \cite{zhou2016making,yu2015statistical,lu2020efficient} 
\begin{equation}
\begin{aligned}
\label{q11_L}
&q_{\alpha}^L=\min: \left( S_+ -  S_-\right) /\big[a_1^\omega a_1^\nu (b_1^\omega b_2^\nu - b_1^\nu b_2^\omega )\big],\\
s.t.:& \\
&\mathcal{F}^-(\hat{n}_\alpha^{lr}) \leq   \hat{n}_\alpha^{lr}  \leq \mathcal{F}^+(\hat{n}_\alpha^{lr}),l,r \in\{\nu,\omega,o\};\\
&\mathcal{F}^-(\hat{n}_\alpha^{l_1r_1}+\hat{n}_\alpha^{l_2r_2}) \leq n_\alpha^{l_1r_1}+n_\alpha^{l_2r_2}  ,l_1r_1,l_2r_2 \in\{\omega\omega,o\nu,\nu o,oo\} \quad and \quad l_1r_1 \neq l_2r_2; \\
&\mathcal{F}^+(\hat{n}_\alpha^{l_1r_1}+\hat{n}_\alpha^{l_2r_2}) \geq   n_\alpha^{l_1r_1}+n_\alpha^{l_2r_2}  ,l_1r_1,l_2r_2 \in\{\nu\nu,o\omega,\omega o\} \quad and \quad l_1r_1 \neq l_2r_2; \\
&\mathcal{F}^-(\hat{n}_\alpha^{l_1r_1}+\hat{n}_\alpha^{l_2r_2}+\hat{n}_\alpha^{l_3r_3}) \leq   n_\alpha^{l_1r_1}+n_\alpha^{l_2r_2}+n_\alpha^{l_3r_3}  ,\\
&l_1r_1,l_2r_2,l_3r_3 \in\{\omega\omega,o\nu,\nu o,oo\} \quad and \quad l_1r_1 \neq l_2r_2 \neq l_3r_3;\\
& \mathcal{F}^+(\hat{n}_\alpha^{l_1r_1}+\hat{n}_\alpha^{l_2r_2}+\hat{n}_\alpha^{l_3r_3}) \geq n_\alpha^{l_1r_1}+n_\alpha^{l_2r_2}+n_\alpha^{l_3r_3}    ,\\
&l_1r_1,l_2r_2,l_3r_3 \in\{\nu\nu,o\omega,\omega o\} \quad and \quad l_1r_1 \neq l_2r_2 \neq l_3r_3;\\
&\mathcal{F}^-(\hat{n}_\alpha^{\omega\omega}+\hat{n}_\alpha^{o \nu }+\hat{n}_\alpha^{\nu o}+n_\alpha^{o o}) \leq   \hat{n}_\alpha^{\omega\omega}+\hat{n}_\alpha^{o \nu }+n_\alpha^{\nu o}+n_\alpha^{o o},\\
&\mathcal{F}^+(\hat{n}_\alpha^{\nu\nu}+\hat{n}_\alpha^{o \omega }+\hat{n}_\alpha^{\omega o}+n_\alpha^{o o}) \geq   \hat{n}_\alpha^{\nu\nu}+\hat{n}_\alpha^{o \omega }+n_\alpha^{\omega o}+n_\alpha^{o o},\\
\end{aligned}
\end{equation}
and
\begin{equation}
\begin{aligned}
\label{q11_U}
&q_{\alpha}^U=\max: (S_+' -  S_-') /  a_1^\omega b_1^\omega,\\
s.t.:& \\
&\mathcal{F}^-(\hat{n}_\alpha^{lr}) \leq   n_\alpha^{lr}  \leq \mathcal{F}^+(\hat{n}_\alpha^{lr}),l,r \in\{\nu,\omega,o\};\\
&\mathcal{F}^-(\hat{n}_\alpha^{o\omega}+\hat{n}_\alpha^{\omega o}) \leq   n_\alpha^{o\omega}+n_\alpha^{\omega o}  ; \\
& \mathcal{F}^+(\hat{n}_\alpha^{\omega\omega}+\hat{n}_\alpha^{oo}) \geq  n_\alpha^{\omega\omega}+n_\alpha^{oo},
\end{aligned}
\end{equation}
where $\hat{n}$ and $\hat{Q}$ are the experimental observable values. The $\mathcal{F}^\pm$ denote the improved Chernoff bound \cite{zhang2017improved} that 
\begin{align}
\mathcal{F}^-(O)=\frac{O}{[1+\delta_1(O)]},\\
\mathcal{F}^+(O)=\frac{O}{[1-\delta_2(O)]},
\end{align}
where $\delta_1(O)$ and $\delta_2(O)$ can be solved by
\begin{align}
\left(\frac{e^{\delta_1}}{(1+\delta_1)^{1+\delta_1}}\right)^\frac{O}{1+\delta_1}=\xi,\\
\left(\frac{e^{-\delta_2}}{(1-\delta_2)^{1-\delta_2}}\right)^\frac{O}{1-\delta_2}=\xi,
\end{align}
where $O$ and $\xi$ denote observed value and the failure probability respectively.

Using the restrictions that
\begin{equation}
\begin{aligned}
&2q^2_{0m}{c_0}^2 + 2q^2_{1m}{c_1}^2 = q_{2m}^2 + q_{3m}^2,\\
&2q^2_{n0}{c'_0}^2 + 2q^2_{n1}{c'_1}^2 = q_{n2}^2 + q_{n3}^2,\\
&{c_0}^2 + {c_1}^2 = 1,\ {c'_0}^2 + {c'_1}^2 = 1,\\
\end{aligned}
\end{equation}
the upper bound of phase error can be expressed as
\begin{equation}
\begin{aligned}\label{The upper bound of phase error}
&e_p^U= \max: \frac{1}{2} - \frac{\left(q_{T_d}-q_{T_s}\right)(q_{01}q_{10}-q_{00}q_{11})^2}{16q_C\sqrt{(q_{10}q_{T1}-q_{11}q_{T0})(q_{00}q_{T1}-q_{01}q_{T0})(q_{01}q_{1T}-q_{11}q_{0T})(q_{00}q_{1T}-q_{10}q_{0T})}   }\\
s.t.&\\
&q_{\alpha}^L \leq q_\alpha \leq q_{\alpha}^U,\\
&\frac{q_{10}q_{T1}-q_{11}q_{T0}}{q_{01}q_{10}-q_{00}q_{11}}+\frac{q_{00}q_{T1}-q_{01}q_{T0}}{q_{11}q_{00}-q_{01}q_{10}}=1, \\
&\frac{q_{01}q_{1T}-q_{11}q_{0T}}{q_{01}q_{10}-q_{00}q_{11}}  +\frac{q_{00}q_{1T}-q_{10}q_{0T}}{q_{11}q_{00}-q_{01}q_{10}}=1.
\end{aligned}
\end{equation}
The target function can also be expressed as
\begin{equation}
\begin{aligned}\label{The upper bound of phase error 1}
&e_p^U= \frac{1}{2} - \frac{\left(q_{T_d}^L-q_{T_s}^U\right)}{16 q_C^U}
\min:\left[\frac{(q_{01}q_{10}-q_{00}q_{11})^2}{\sqrt{(q_{10}q_{T1}-q_{11}q_{T0})(q_{00}q_{T1}-q_{01}q_{T0})(q_{01}q_{1T}-q_{11}q_{0T})(q_{00}q_{1T}-q_{10}q_{0T})}}\right].
\end{aligned}
\end{equation}
In other words, we only need to estimate the lower bound of 
\begin{equation}
\begin{aligned}\label{lower bound 1}
\frac{(q_{01}q_{10}-q_{00}q_{11})^2}{\sqrt{(q_{10}q_{T1}-q_{11}q_{T0})(q_{00}q_{T1}-q_{01}q_{T0})(q_{01}q_{1T}-q_{11}q_{0T})(q_{00}q_{1T}-q_{10}q_{0T})}}.
\end{aligned}
\end{equation}
We define 
\begin{equation}
\begin{aligned}
\label{Intermediate variables X and Y}
\mathcal{X}=\sqrt{\frac{q_{01}q_{T0}-q_{00}q_{T1}}{q_{10}q_{T1}-q_{11}q_{T0}}},
\mathcal{Y}=\sqrt{\frac{q_{10}q_{0T}-q_{00}q_{1T}}{q_{01}q_{1T}-q_{11}q_{0T}}}.
\end{aligned}
\end{equation}
By employing the intermediate variables $\mathcal{X}$ and $\mathcal{Y}$ and the constraints
\begin{equation}
\begin{aligned}
& c_0^2 + c_1^2 = \frac{q_{10}q_{T1}-q_{11}q_{T0}}{q_{01}q_{10}-q_{00}q_{11}}+\frac{q_{00}q_{T1}-q_{01}q_{T0}}{q_{11}q_{00}-q_{01}q_{10}}=1, \\
& {c'_0}^2 + {c'_1}^2 =\frac{q_{01}q_{1T}-q_{11}q_{0T}}{q_{01}q_{10}-q_{00}q_{11}}  +\frac{q_{00}q_{1T}-q_{10}q_{0T}}{q_{11}q_{00}-q_{01}q_{10}}=1,
\end{aligned}
\end{equation}
Eq.(\ref{lower bound 1}) can be rewritten as
\begin{equation}
\left(\mathcal{X}+\frac{1}{\mathcal{X}}\right)\times\left(\mathcal{Y}+\frac{1}{\mathcal{Y}}\right),
\end{equation}
whose lower bound can be expressed as 
\begin{equation}
\begin{aligned}
\left(\mathcal{X}+\frac{1}{\mathcal{X}}\right)^L= \left \{
\begin{array}{lc}
\mathcal{X}^L+1/{\mathcal{X}^L}, &\mathcal{X}^L>1\\
\mathcal{X}^U+1/{\mathcal{X}^U},&\mathcal{X}^U<1\\
2,&else 
\end{array}
\right.\\
\left(\mathcal{Y}+\frac{1}{\mathcal{Y}}\right)^L= \left \{
\begin{array}{lc}
\mathcal{Y}^L+1/{\mathcal{Y}^L}, &\mathcal{Y}^L>1\\
\mathcal{Y}^U+1/{\mathcal{Y}^U},&\mathcal{Y}^U<1\\
2,&else 
\end{array}
\right.
\end{aligned}
\end{equation}
where
\begin{equation}
\begin{aligned}
\mathcal{X}^L=\sqrt{\frac{q_{01}^L q_{T0}^L -q_{00}^U q_{T1}^U}{q_{10}^U q_{T1}^U -q_{11}^L q_{T0}^L}},
\mathcal{X}^U=\sqrt{\frac{q_{01}^U q_{T0}^U -q_{00}^L q_{T1}^L}{q_{10}^L q_{T1}^L -q_{11}^U q_{T0}^U}},\\
\mathcal{Y}_L=\sqrt{\frac{q_{10}^L q_{0T}^L -q_{00}^U q_{1T}^U}{q_{01}^U q_{1T}^U -q_{11}^L q_{0T}^L}},
\mathcal{Y}_U=\sqrt{\frac{q_{10}^U q_{0T}^U -q_{00}^L q_{1T}^L}{q_{01}^L q_{1T}^L -q_{11}^U q_{0T}^U}}.
\end{aligned}
\end{equation}

The secret key rate is 
\begin{equation}
\begin{aligned}
R = p_\mu p_{\mu'}\left[ a_1^\mu b_1^{\mu} {q}^L_{C}\left( 1 - H({e_p^U}) \right) - Q_{C}^{\mu\mu}fH(E_{C}^{\mu\mu}) \right],
\end{aligned}
\end{equation}
where $p_\mu$ ($p_{\mu'}$) denote the probability of Alice (Bob) selecting intensity code basis with intensity $\mu$ ($\mu'$).

To demonstrate the maximal tolerated imbalance, we simulated our decoy-state protocol with different fiber lengths and different data sizes. As illustrated in Fig.\ref{fig_misalignment_tolerance_pure}, we can find that its property of the imbalance tolerance increases with the data size. In non-asymptotic cases, the protocol performance slowly decreases with the increasing $\beta$ and in asymptotic cases, our protocol has nearly invariant performance.
\begin{figure}[htbp]
 \subfigure[Fiber length = 25 km]{	
\includegraphics[width=7.9cm]{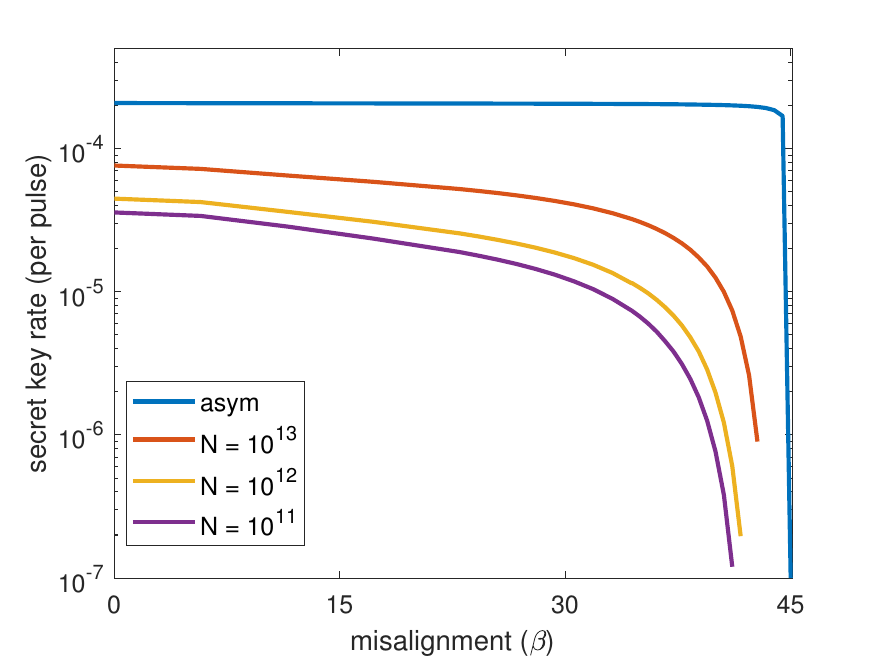} 
}
 \subfigure[Fiber length = 50 km]{	
\includegraphics[width=7.9cm]{pure_state_R_theta_L=50.pdf} 
}
\caption{\label{fig_misalignment_tolerance_pure} Secret key rate as a function of the imbalance $\beta$ ($\beta' = \beta$) in the pure-state scenarios. The blue lines denote the asymptotic case that the data size is infinite. The red, yellow, and purple lines denote the secret key rates with the data size of $10^{13}$, $10^{12}$, $10^{11}$ respectively.}
\end{figure}

\subsection{Mixed-state scenario }

The decoy-state analysis for the mixed-state scenario is a little different from the pure-state scenario because we should bound the mixed-state yields $y_{\alpha}$ first and estimate the $q_{\alpha}$ by the upper and lower bound of the $y_{\alpha}$, where $\alpha\in\{00,01,10,11,T_s,T_d,T0,T1,0T,1T,C\}$.

The upper and lower bound of $y_{\alpha}$ can be estimated by the equations similar to Eq.(\ref{q11}), Eq.(\ref{Eq_S}), Eq.(\ref{q11_L}), and Eq.(\ref{q11_U}) except that all pure-state single-photon yields "$q_\alpha$" should be replaced by mixed-state single-photon yields "$y_\alpha$". The $q_\alpha$ can be bounded by equations similar to Eq.(\ref{q_nm bounds ZX}) and Eq.(\ref{q_nm bounds ZZ}) except that the $y_\alpha$ are replaced by its upper or lower bound for the worst-case estimation:
\begin{equation}
\begin{aligned}
&q_{00}^L=\frac{(1-\zeta^U)(1-\zeta'^U)y_{00}^L-(1-\zeta^L)\xi^Uy_{01}^U-\xi^U(1-\zeta'^L)y_{10}^U+\xi^L\xi'^L y_{11}^L}{(1-\xi^L-\zeta^L)(1-\xi'^L-\zeta'^L)},\\
&q_{00}^U=\frac{(1-\zeta^L)(1-\zeta'^L)y_{00}^U-(1-\zeta^U)\xi^Ly_{01}^L-\xi^L(1-\zeta'^U)y_{10}^L+\xi^U\xi'^U y_{11}^U}{(1-\xi^U-\zeta^U)(1-\xi'^U-\zeta'^U)},\\
&q_{01}^L=\frac{-(1-\zeta^L)\zeta'^Uy_{00}^U+(1-\zeta^U)(1-\xi'^U)y_{01}^L+\xi^L\zeta'^Ly_{10}^L-\xi^U(1-\xi'^L)y_{11}^U}{(1-\xi^L-\zeta^L)(1-\xi'^L-\zeta'^L)},\\
&q_{01}^U=\frac{-(1-\zeta^U)\zeta'^Ly_{00}^L+(1-\zeta^L)(1-\xi'^L)y_{01}^U+\xi^U\zeta'^Uy_{10}^U-\xi^L(1-\xi'^U)y_{11}^L}{(1-\xi^U-\zeta^U)(1-\xi'^U-\zeta'^U)},\\
&q_{10}^L=\frac{-\zeta^U(1-\zeta'^L)y_{00}^U+\zeta^L\xi'^Ly_{01}^L+(1-\xi^U)(1-\zeta'^U)y_{10}^L-(1-\xi^L)\xi'^Uy_{11}^U}{(1-\xi^L-\zeta^L)(1-\xi'^L-\zeta'^L)},\\
&q_{10}^U=\frac{-\zeta^L(1-\zeta'^U)y_{00}^L+\zeta^U\xi'^Uy_{01}^U+(1-\xi^L)(1-\zeta'^L)y_{10}^U-(1-\xi^U)\xi'^Ly_{11}^L}{(1-\xi^U-\zeta^U)(1-\xi'^U-\zeta'^U)},\\
&q_{11}^L=\frac{\zeta^L\zeta'^Ly_{00}^L-\zeta^U(1-\xi'^L)y_{01}^U-(1-\xi^L)\zeta'^Uy_{10}^U+(1-\xi^U)(1-\xi'^U)y_{11}^L}{(1-\xi^L-\zeta^L)(1-\xi'^L-\zeta'^L)},\\
&q_{11}^U=\frac{\zeta^U\zeta'^Uy_{00}^U-\zeta^L(1-\xi'^U)y_{01}^L-(1-\xi^U)\zeta'^Ly_{10}^L+(1-\xi^L)(1-\xi'^L)y_{11}^U}{(1-\xi^U-\zeta^U)(1-\xi'^U-\zeta'^U)},\\
\end{aligned}
\end{equation}
\begin{equation}
\begin{aligned}
&q_{T0}^L=\frac{(1-\zeta'^U)y_{T0}^L-\xi'^Uy_{T1}^U}{1-\xi'^U-\zeta'^U},\ \
q_{T0}^U=\frac{(1-\zeta'^L)y_{T0}^U-\xi'^Ly_{T1}^L}{1-\xi'^L-\zeta'^L},\\
&q_{T1}^L=\frac{(1-\xi'^U)y_{T1}^L-\zeta'^Uy_{T0}^U}{1-\xi'^U-\zeta'^U},\ \
q_{T1}^U=\frac{(1-\xi'^L)y_{T1}^U-\zeta'^Ly_{T0}^L}{1-\xi'^L-\zeta'^L},\\
&q_{0T}^L=\frac{(1-\zeta^U)y_{0T}^L-\xi^Uy_{1T}^U}{1-\xi^U-\zeta^U},\ \
q_{0T}^U=\frac{(1-\zeta^L)y_{0T}^U-\xi^Ly_{1T}^L}{1-\xi^L-\zeta^L},\\
&q_{1T}^L=\frac{(1-\xi^U)y_{T1}^L-\zeta^Uy_{T0}^U}{1-\xi^U-\zeta^U},\ \
q_{1T}^U=\frac{(1-\xi^L)y_{T1}^U-\zeta^Ly_{T0}^L}{1-\xi^L-\zeta^L},\\
&q_{Ts}^L=y_{Ts}^L,\ \
q_{Ts}^U=y_{Ts}^U,\ \
q_{Td}^L=y_{Td}^L,\ \
q_{Td}^U=y_{Td}^U,\\ 
\end{aligned}
\end{equation}

As long as the $q_\alpha$ are bounded, the $e_p$ can be estimated by the same method as the pure-state scenario. The final secret key rate is expressed as 
\begin{equation}
\begin{aligned}
R = p_\mu p_{\mu'}\left[ a_1^\mu b_1^{\mu} {y}^L_{C}\left( 1 - H({e_p^U}) \right) - Q_{C}^{\mu\mu}fH(E_{C}^{\mu\mu}) \right].
\end{aligned}
\end{equation}

We also simulated the performance and the imbalance tolerance for the mixed-state scenario. The results are illustrated in Fig.\ref{fig_misalignment_tolerance_mixed}, which indicates that our protocol also have a good imbalance tolerance in the mixed-state scenarios.

\begin{figure}[htbp]
 \subfigure[Fiber length = 25 km]{	
\includegraphics[width=7.9cm]{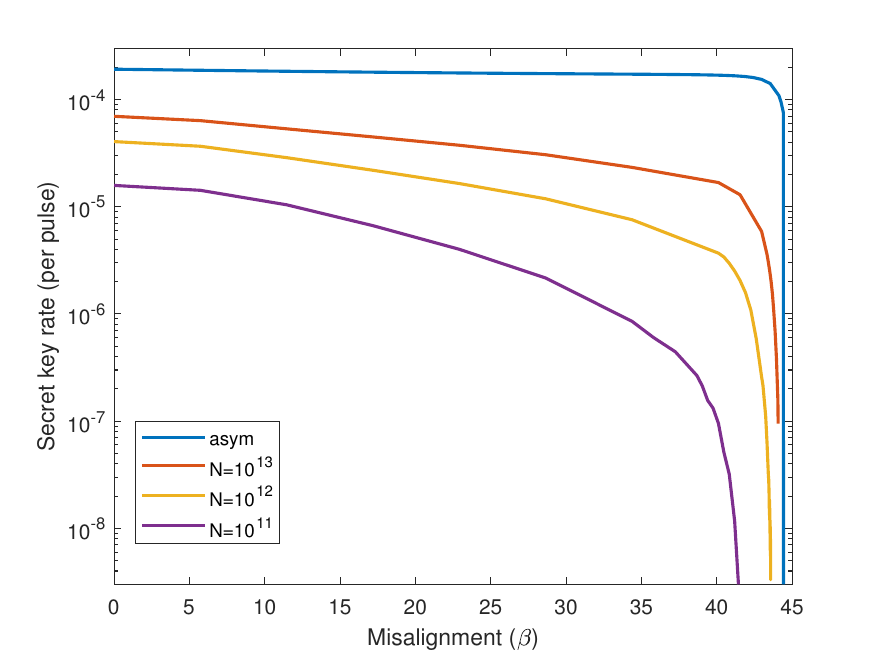} 
}
 \subfigure[Fiber length = 50 km]{	
\includegraphics[width=7.9cm]{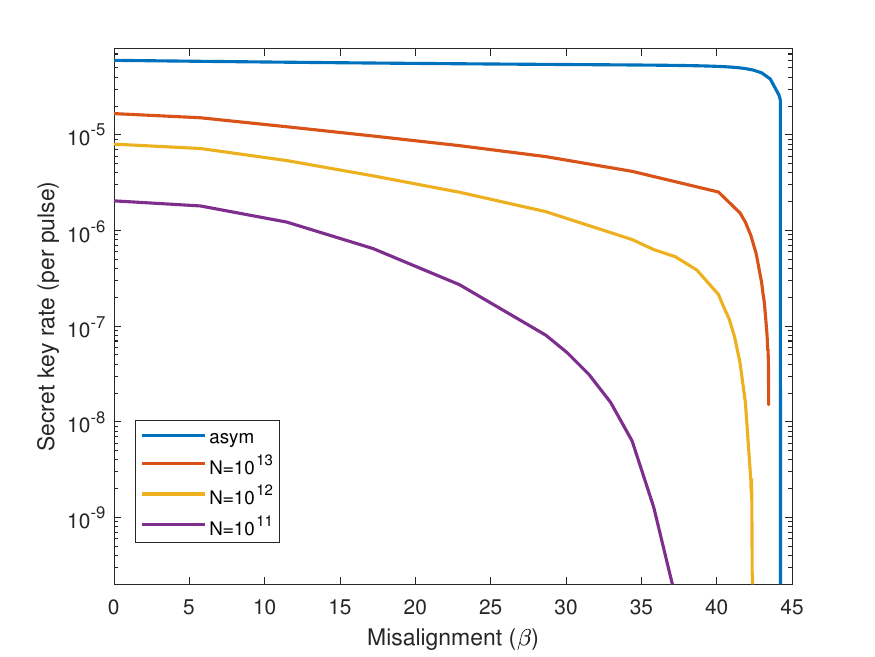} 
}
\caption{\label{fig_misalignment_tolerance_mixed} Secret key rate as a function of the imbalance $\beta$ ($\beta' = \beta$) in the mixed-state scenarios. The blue lines denote the asymptotic case that the data size is infinite. The red, yellow, and purple lines denote the secret key rates with the data size of $10^{13}$, $10^{12}$, $10^{11}$ respectively.}
\end{figure}


\section{Parameter optimization}

For the best performance, Alice and Bob set same experimental parameters and optimize \cite{xu2013practical} eight of them, including three intensities $\mu$, $\nu$, $\omega$, three probabilities $p_\mu$, $p_\nu$, $p_\omega$ for selecting the corresponding intensity; and two conditional probabilities $p_{C|\nu}$, $p_{C|\omega}$ for selecting the code basis on the condition that the corresponding intensity has been selected. As listed in Tab. \ref{Tab_experimental_parameters}, three sets of parameters are optimized for the three different cases respectively.
The remaining parameters are constrained by relations as follows: the probability of selecting $o$ is $p_o = 1 - p_\mu - p_\nu - p_\omega$; the probabilities of selecting the test basis are $p_{T|\nu} = 1 - p_{C|\nu}$ and $p_{T|\omega} = 1 - p_{C|\omega}$. The experimental data is processed by the two different post-processing methods, namely, the pure-state method and the mixed-state method. The results and the important intermediate variables for the two scenarios are listed in Tab.\ref{Tab_experimental_results1} and Tab.\ref{Tab_experimental_results2} respectively.

\begin{table}[htbp]
\caption{\label{Tab_experimental_parameters}
Optimized decoy-state method parameters of the experiment.}
\begin{ruledtabular}
\begin{tabular}{ccc|cccccccc}
 & $\beta$ & $\beta'$ & $\mu$& $\nu$& $\omega$& $p_\mu$& $p_\nu$& $p_\omega$ & $p_{C|\nu}$ & $p_{C|\omega}$  \\
\hline
\textit{case 1} & 0 & 0 & $0.349$& $0.239$& $0.0515$& $0.463$& $0.1$& $0.357$ & $0.412$ & $0.391$  \\
\textit{case 2} & 0 & $10^{\circ}$ & $0.312$& $0.231$& $0.0509$& $0.437$& $0.1$& $0.380$ & $0.592$ & $0.384$  \\
\textit{case 3} & $10^{\circ}$ & $10^{\circ}$ & $0.295$& $0.237$& $0.0509$& $0.411$& $0.1$& $0.400$ & $0.423$ & $0.395$  \\
\end{tabular}
\end{ruledtabular}
\end{table}

\begin{table}[htbp]
\caption{\label{Tab_experimental_results1}
Experimental results of the pure-state scenarios}
\begin{ruledtabular}
\begin{tabular}{ccc|ccccc}
imbalances & $\beta$ & $\beta'$ & $Q_C^{\mu\mu}$& $E_C^{\mu\mu}$& $\underline{e}_p$ & $\underline{q}_C$ & $R$  \\
\hline
\textit{case 1} & 0 & 0 & 5.98E-5  & $0.0086$ & $0.166$& 4.71E-4 & 1.10E{-6}\\
\textit{case 2} & 0 & $10^{\circ}$ & 4.77E-5  & $0.0091$ & $0.189$& 5.09E-4 & 7.37E{-7}\\
\textit{case 3} & $10^{\circ}$ & $10^{\circ}$ & 4.29E-5  & $0.0097$ & $0.189$& 5.08E-4 & 5.87E{-7}\\
\end{tabular}
\end{ruledtabular}
\end{table}

\begin{table}[htbp]
\caption{\label{Tab_experimental_results2}
Experimental results of the mixed-state scenarios}
\begin{ruledtabular}
\begin{tabular}{ccc|ccccc}
 imbalances & $\beta$ & $\beta'$ & $Q_C^{\mu\mu}$& $E_C^{\mu\mu}$& $\underline{e}_p$ & $\underline{y}_C$ & $R$  \\
\hline
\textit{case 1} & 0 & 0 & 5.98E-5  & $0.0086$ & $0.166$& 4.71E-4 & 1.10E{-6}\\
\textit{case 2} & 0 & $10^{\circ}$ & 4.77E-5  & $0.0091$ & $0.189$& 5.09E-4 & 7.37E{-7}\\
\textit{case 3} & $10^{\circ}$ & $10^{\circ}$ & 4.29E-5  & $0.0097$ & $0.189$& 5.08E-4 & 5.87E{-7}\\
\end{tabular}
\end{ruledtabular}
\end{table}

\section{Discussion}

In summary, we have proposed a MDIQKD that has the same performance compared with the original MDIQKD while fewer assumptions in encoding systems are required. The new protocol has higher security since it is not only measurement device independent, but also immune to some side-channel attacks stems from the imperfect coding. In addition, it is more experimentally convenient since it simplifies the encoding system. We have successfully validated the protocol with a practical MDIQKD system, which is automatically calibrated and works continuously to collect sufficient data. We have collected $5\times 10^{11}$ pulse pairs in each of the three different imbalances and successfully generate positive key rate. The performance would be further improved by employing higher clockwork frequency, superconducting nanowire single-photon detectors, and improved decoy-state methods. Due to the higher security, simpler coding method, and comparable key rate, this protocol would be a benefit for practical applications, especially in the scenarios that the precision of control systems is limited or the calibration is difficult, such as the network scenarios. The performance in non-asymmetric scenarios could be further improved by designing more efficient decoy-state methods or introducing better analysis for the finite-key size effect. This work would also provide inspiration for designing protocols with fewer assumptions by combining with other protocols.

\section{Funding.}
This work was supported by the National Key Research And Development Program of China (Grant No. 2018YFA0306400), the National Natural Science Foundation of China (Grants Nos. 61961136004,62171424,621714418). Rong Wang is financially supported by the University of Hong Kong start-up grant.
\section{Disclosures.}
The authors declare no conflicts of interest.

\section{Data Availability.}
Data underlying the results presented in this paper are not publicly available at this time but may be obtained from the authors upon reasonable request.

\end{widetext}

\section{REFERENCES}
\bibliography{citations}
\end{document}